\UseRawInputEncoding

\documentclass[a4paper,11pt]{article}

\usepackage{jcappub} 

\usepackage[T1]{fontenc} 
\usepackage{subfigure}

\title{\boldmath Towards high temperature holographic superconductors}

\author[a,b,1]{Mahya Mohammadi,\note{Corresponding author.}}
\author[a,b]{Ahmad Sheykhi}

\affiliation[a]{Department of Physics, 
College of Sciences, Shiraz University, Shiraz 71454, Iran}
\affiliation[b]{Biruni Observatory, College of Sciences, Shiraz 
University, Shiraz 71454, Iran}

\emailAdd{mahya.mohammadi@hafez.shirazu.ac.ir}
\emailAdd{asheykhi@shirazu.ac.ir}

\abstract{We explore a holographic superconductor model in which a real 
scalar field is non-minimally coupled to a gauge field. We
consider several types of the non-minimal coupling function
$h(\psi)$ including exponential, hyperbolic (cosh), power-law and
fractional forms. We investigate the influences of the non-minimal
coupling parameter $\alpha$ on condensation, critical temperature
and conductivity. We can categorize our results in two groups. In
the first group, conductor/superconductor phase transition is
easier to occur for larger values of $\alpha$, while in the second
group stronger effects of the non-minimal coupling makes the
formation of scalar hair harder. Although the real and imaginary
parts of conductivity are impressed by different forms of
$h(\psi)$, they follow some universal behaviors such as connecting
with each other through Kramers-Kronig relation in
$\omega\rightarrow 0$ limit or the appearance of gap frequency at
low temperatures around $\omega_g\sim8 T_c$ which shifts to larger
values by increasing the strength of $\alpha$. Among all forms of
$h(\psi)$ we observe that $h(\psi)=1+\alpha \psi^2$ gives us
better information in wide range of non-minimal coupling constant
and temperature. Choosing the best form of $h(\psi)$, we construct
a family of solutions for holographic conductor/superconductor
phase transitions to discover the effect of the hyperscaling
violation when the gauge and scalar fields are non-minimally
coupled. we find that the critical temperature increases for
higher effects of hyperscaling violation $\theta$ and non-minimal
coupling constant $\alpha$. By increasing these two parameters, we
obtain lower values of condensation which means that
conductor/superconductor phase transition will acquire easier.
Furthermore, we understand that the hyperscaling violation affects
the conductivity $\sigma$ of the holographic superconductors and
changes the expected relation in the gap frequency. Some universal
behaviors like infinite DC conductivity are observed.}


\begin{document}
\maketitle
\flushbottom

\section{Introduction}
Nowadays, we are able to examine the validity of General 
Relativity (GR) by using observational evidences coming from
gravitational waves detections and the first image of the black
hole shadow using Event Horizon Telescope (EHT) collaboration
\cite{01,02,03,04}. In spite of achieving a lot of success in
explaining classical gravitational phenomena, GR is inadequate to
explain several problems at the interface between gravitation and
high energy physics. In this regards, the physics beyond GR has
been arisen to investigate classical and quantum gravity. The
motivation is to improve our understanding in different aspects of
gravity such as explaining the late-time acceleration of the
universe, space-time singularities, dark matter, gravitational
waves and inflation epoch \cite{01,02,03,04}. Spontaneous
scalarization is a famous notion beyond GR theory. Black hole
scalarization is one of the key concepts of holographic
superconductors theory due to the spontaneous U(1) symmetry
breaking which yields to a phase transition from a black hole with
no hair to a hairy one for temperatures below the critical value
\cite{H08}.

Holographic superconductors are constructed based on AdS/CFT
duality which is a powerful tool to study strongly correlated
systems that relates a weak gravitational system in $(d +
1)$-dimensions to the strong coupling conformal field theory in
$d$-dimensions \cite{Maldacena,H08,G98,W98,HR08,R10,H11,H09}. The
idea of holographic superconductors has attracted a lot of
attentions as a theoretical insight to real high temperature
superconductors by showing similar behaviors as real
supercondutors like second order phase transition
\cite{Hg09,Gu09,HHH08,JCH10,cai15,SH16,SHsh(17),
Ge10,Ge12,Kuang13,Pan11,CAI11, SHSH(16),shSh(16),
cai10,yao13,n4,n5,n6,Gan1,mahya,BCS57,superp,Caip,cai13p,Donos,Gubser,chaturverdip15,Roberts8,
zeng11,cai11p,pando12,momeni12p,mahyap,francessco1,francessco2,mahyalif,mahyap11,mahhya,mahhhya}.
There are still many open problems in strongly coupled systems
that each of them can help us to understand the physics of
condensed matter systems in this regime. It is worth investigating
the effects of the non-minimal coupling on the holographic
condensed matter systems. General models for holographic
superconductivity parameterized by three couplings which are
functions of a real scalar field \cite{070}. It was confirmed that
some features are universal and independent of the model such as
DC conductivity but important features of the quantum critical
behavior strongly depend on the choice of couplings, such as the
order of the phase transition and critical exponents of the
second-order phase transitions.

In the present work, we concentrate on a modified GR theory dubbed
Einstein-Maxwell-Scalar (EMS) theory. We suppose a real scalar
field is non-minimally coupled to a gauge field through different
forms of coupling functions $h(\psi)$ \cite{05,06,07}. Effects of
exponential form of the non-minimal coupling function on the phase
transition of the holographic superconductors in EMS theory were
addressed \cite{07}. Here, we extend the work of \cite{07} by
considering different forms of $h(\psi)$ to check how the results
depend on the non-minimal coupling function. We solve equations of
motion numerically to analyze the influence of the non-minimal
coefficient $\alpha$ in each kind of coupling function $h(\psi)$
on the critical temperature $T_c$. Moreover in \cite{070}, a phase
transition from the second order to the first order appeared in
some conditions. Therefore, we explore phase transition of the
system to check whether or not different values of $\alpha$ in
each model of coupling will change the order of phase transition.
In \cite{05}, spontaneous scalarization for the negative values of
the non-minimal coupling term $\alpha$ were reported. However,
since in holographic setup the formation of scalar hair appears by
going down the temperatures below the critical value, we consider
positive values of $\alpha$, too. By applying an appropriate
electromagnetic perturbation on the black hole background, we
explore the behavior of the real and imaginary parts of
conductivity for different values of $\alpha$ in each form of
$h(\psi)$. We would like to explore how the universal relation
between the gap frequency and critical temperature in holographic
setup, $\omega_g\sim 8 T_c$, can be changed based on different
choices of the non-minimal component.

Asymptotically AdS geometry corresponds to relativistically
conformal theory in the boundary but the condensed matter theory
is non-relativistic in most cases. Therefore, it is interesting to
generalize the holographic superconductor models to
nonrelativistic situations by using the nonrelativistic version of
AdS/CFT correspondence. This is achieved in Lifshitz-like geometry
with dynamical exponent in which the time scales different from
the space, which is similar to the behavior of the condensed
matter systems near the quantum critical point and brings some new
features of holographic superconductor models (see e.g.
\cite{LU,Natsuume,Sherkatghanad,Zhao14,lif1,lif2,lif3} and
references therein). On the other side, hyperscaling violation
metric, as an extension of the Lifshitz metric, has grabbed a lot
of attentions because of its potential to describe holographically
the realistic condensed matter systems
\cite{hyper1,hyper2,hyper3,hyper4,hyper5,hyper6}. We shall also
construct solutions of the holographic superconductors in the
background of the hyperscaling violation black hole when gauge and
scalar fields are non-minimally coupled. We employ the numerical
method to analyze the influences of the hyperscaling violating
parameter, $\theta$, and the non-minimal coupling parameter,
$\alpha$, on the condensation, critical temperature and
conductivity of the system. In \cite{hyper4}, it was mentioned
that by increasing $\theta$, the critical temperature increases
analytically but first decreases and then increases numerically.
Therefore, the behavior of the critical temperature $T_c$ due to
various values of $\alpha$ and $\theta$ should be checked.
Moreover, the second order phase transition is one of main
concepts of superconductors. We will check whether or not the
order of phase transition will differ by applying different
strength of $\alpha$ and $\theta$. In addition, we will study the
electrical conductivity in holographic setup to check the
universal behaviors of conductivity like infinite Dc conductivity
and $\omega_g \sim 8 T_c$.

This paper is organized as follows. In section \ref{sec2}, we
introduce the holographic model when scalar and gauge fields are
non-minimally coupled via different forms of the coupling
functions. Then, we analyze condensation and critical temperature
of $(3+1)$-dimensional holographic superconductor impressed by
various strength of non-minimal coupling constant $\alpha$ for
each model. After that we investigate the role of the non-minimal
coupling on the real and imaginary parts of conductivity. Next in
\ref{sec2.2}, we present the holographic setup with hyperscaling
violation by studying the behavior of condensation, critical
temperature and conductivity. Finally in section \ref{sec3}, we
summarize our results.

\section{Holographic Setup}\label{sec2} 
The action of the four dimensional gravity in the presence of a
real scalar field $\psi$ with mass $m$ and charge $q$ which is
non-minimally coupled to the Maxwell gauge field is given by
\cite{05,06,07},
\begin{equation}\label{eqact}
S =\int d^{4}x\sqrt{-g} \left[ R+\frac{6}{L^{2}}-\frac{h(\psi)}{4} F_{\mu\nu}F^{\mu\nu}- \vert\nabla\psi-i q A \psi \vert^{2}
-m^{2} \vert \psi \vert^{2}\right],
\end{equation}
where $g$, $R$ and $L$ represent metric determinant, Ricci scalar
and the curvature radius of AdS spacetime, respectively. In the
above action $A_{\mu}$ stands for the gauge field that connects to
the electromagnetic field strength tensor $F_{\mu\nu}$ through
$F_{\mu\nu}=\partial_{\mu}A_{\nu}-\partial_{\nu}A_{\mu}$.
$h(\psi)$ denotes the non-minimal coupling of the scalar field and
Maxwell background which has various coupling choices. In this
work we will consider some forms of non-minimal coupling function
$h(\psi)$ with $\alpha$ as the coupling constant as below
\cite{05,06,07}

\[
h(\psi) =
\begin{cases}
e^{\pm\alpha \psi^{2}} & \text{Exponential Coupling (Exp$(\pm)$)},\\
cosh(\sqrt{2 \vert \alpha \vert}\psi) & \text{Hyperbolic Cosine Coupling (HCos)},\\
1\pm\alpha \psi^{2} & \text{Power Coupling (P$(\pm)$)},\\
\dfrac{1}{1+\alpha \psi^{2}} & \text{Fractional Coupling (F)},\\
\end{cases}
\]
for $\alpha \psi^{2}\ll 1$, we have
\begin{eqnarray}
&& e^{-\alpha \psi^2} \simeq \dfrac{1}{1+\alpha \psi^2} \simeq 1-\alpha \psi^2 , \label{ex1}\\
&& cosh(\sqrt{|2 \alpha|} \psi) \simeq e^{\alpha \psi^2} \simeq 1+\alpha \psi^2 \label{ex2}.
\end{eqnarray}
In the probe limit when the scalar and electromagnetic fields do not backreact, the metric of a planar Schwarzschild-AdS black hole has the following form \cite{H08}
\begin{eqnarray} \label{metric2}
&&{ds}^{2}=-f(r){dt}^{2}+\frac{{dr}^{2}}{f(r)}+r^{2}({dx}^{2}+{dy}^{2})%
,\\
&&f(r)=\frac{r^2}{L^{2}} (1-\frac{r_+^{3}}{r^{3}})\label{eqf}, %
\end{eqnarray}%
with $r_+$ as the horizon location. The Hawking temperature of the black hole is given by \cite{H08}
\begin{equation}
T=\frac{f'(r_+)}{4 \pi}=\frac{3 r_+}{4 \pi L^{2}},
\end{equation}
without loss of generality, we set $L=r_+=1$ for the rest of this work.
\subsection*{Condensation of the scalar field}
We can obtain the equations of motion by varying the action (\ref{eqact}) with respect to scalar field $\psi$ and gauge field $A_{\mu}$ as
\begin{gather}
\nabla _{\mu } \nabla ^{\mu } \psi -q^{2} A_{\mu } A^{\mu } \psi - m^{2}\psi -\frac{\dot{h}(\psi)}{8} F_{\mu \nu} F^{\mu \nu}=0, \label{Epsi} \\
\notag \\
2 q^{2} A_{\nu} \psi^{2}-\nabla ^{\mu }\left( h(\psi) F_{\mu \nu}\right) =0,
\end{gather}
where $\dot{h}=dh/d\psi$. Choosing $\psi=\psi(r)$ and $A_{\mu}dx^{\mu}=\phi(r)dt$ yields to
\begin{equation}\label{eqpsi}
\psi ''+\left[\frac{2}{r}+\frac{f'}{f}\right] \psi '+ \left[\frac{q^2 \phi ^2}{f^2}-\frac{m^2}{f}\right]\psi+\frac{\phi '^2}{4 f}\dot{h}(\psi)=0,
\end{equation}
\begin{equation}\label{eqphi}
\phi ''+\left[\frac{2}{r}+\frac{h'(\psi)}{h(\psi)}\right] \phi '
-\frac{2 q^2 \psi^2 }{f h(\psi)}\phi =0,
\end{equation}
in which primes denote the derivatives with respect to the radial coordinate $r$.
If we consider $\mu$ and $\rho$ as chemical potential and charge density, the above equations have the asymptotic behavior($r\rightarrow\infty$) as
\begin{eqnarray}
&& \psi=\frac{\psi_{+}}{r^{\Delta _+}}+\frac{\psi _{-}}{r^{\Delta _-}}, \ \ \ \Delta _\pm=\left[3\pm\sqrt{9+ 4 m^2}\right],
\\
&& \phi=\mu -\frac{\rho }{r}.
\end{eqnarray}

\begin{figure*}[t]
\centering
\subfigure[~Exp(-)]{\includegraphics[width=0.4\textwidth]{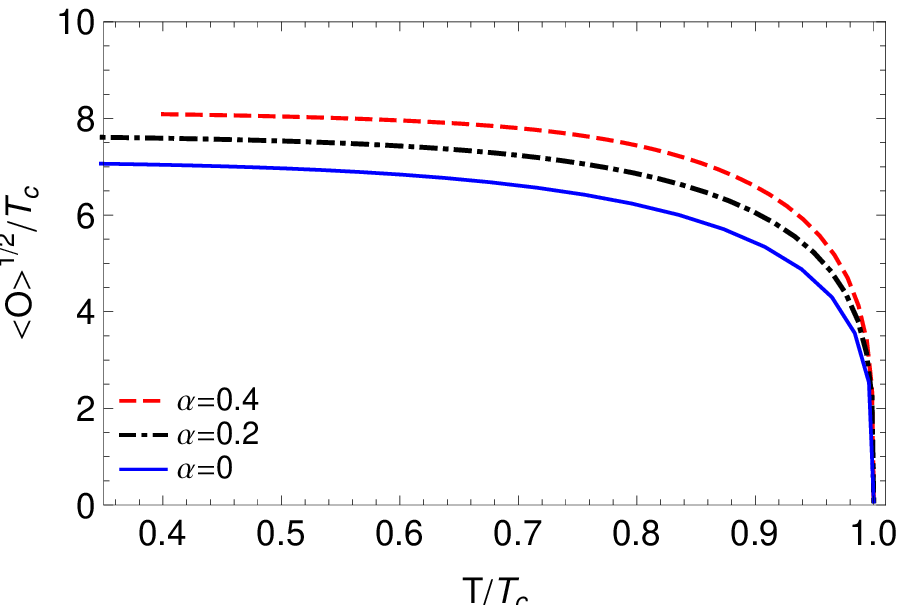}} \qquad %
\subfigure[~Exp(+)]{\includegraphics[width=0.4\textwidth]{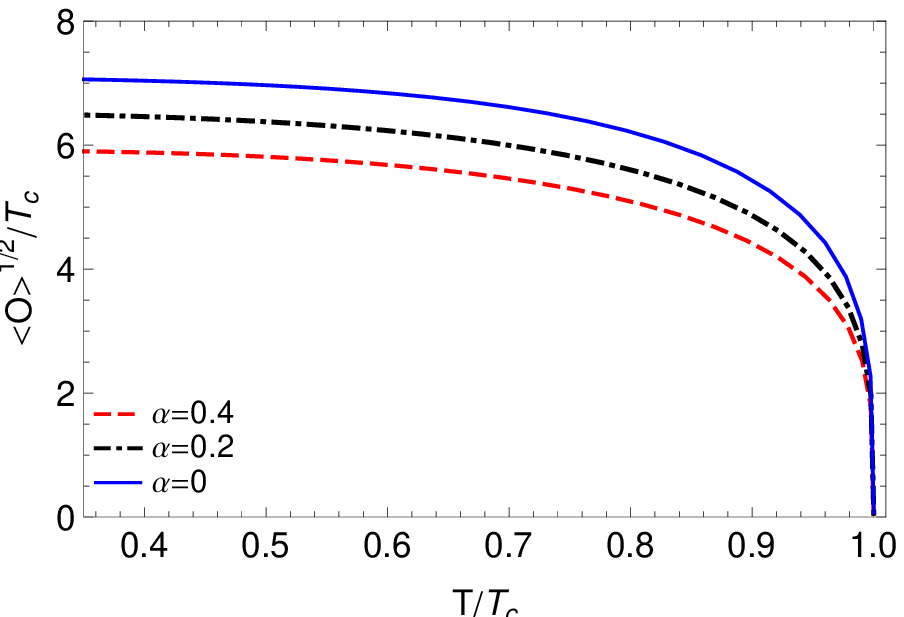}} \qquad %
\subfigure[~F]{\includegraphics[width=0.4\textwidth]{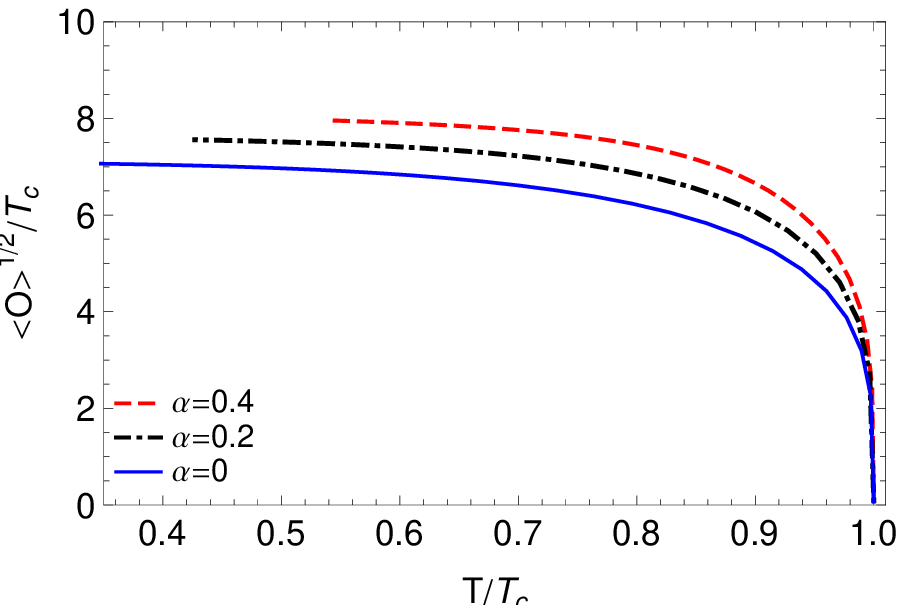}} \qquad %
\subfigure[~HCos]{\includegraphics[width=0.4\textwidth]{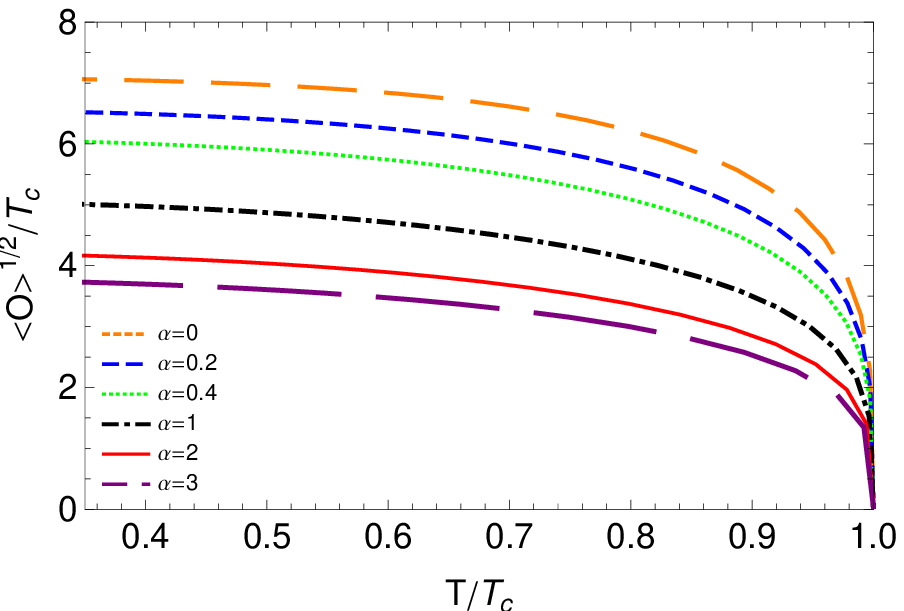}} \qquad %
\subfigure[~P(-)]{\includegraphics[width=0.4\textwidth]{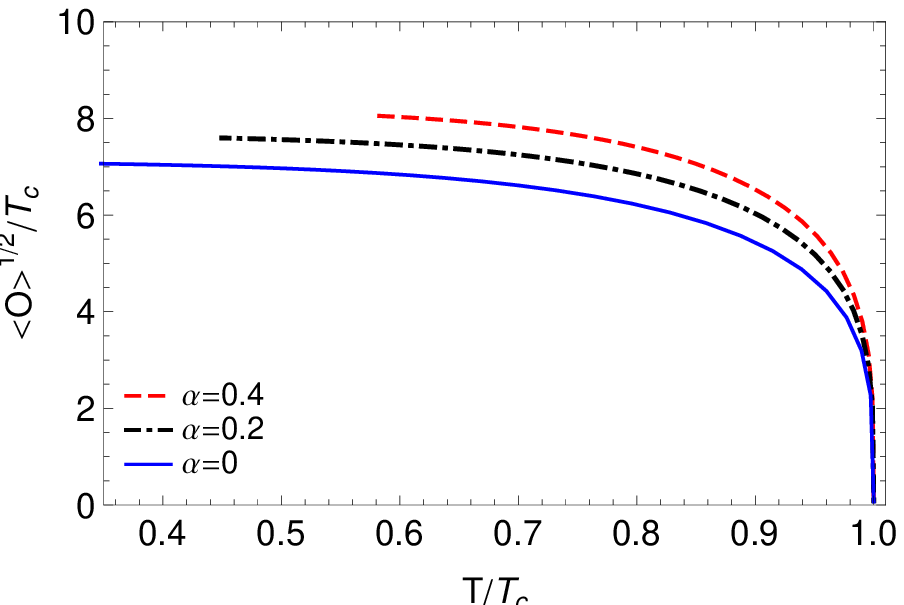}} \qquad %
\subfigure[~P(+)]{\includegraphics[width=0.4\textwidth]{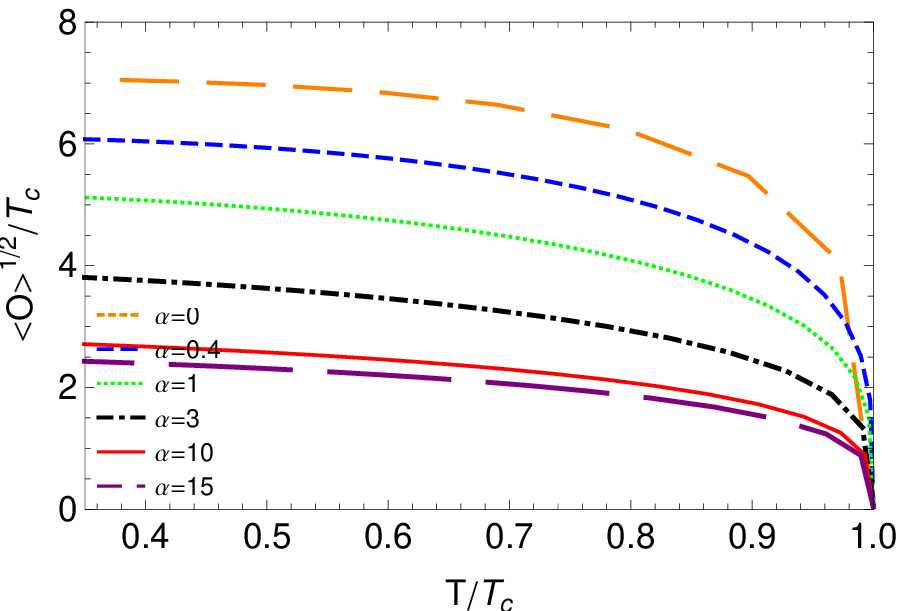}} \qquad %

\caption{The behavior of condensation as a function of temperature.}\label{fig1}
\end{figure*}

In this work, we take $m^2= -2$ and $q=1$. Since we are looking
for spontaneous phase transition, we set $\psi _{-}=0$ which means
the condensate turns on without being sourced. $\psi _{+}$ plays
the role of expectation value of the order parameter $\langle
O\rangle$. The effect of different values of coupling constant
$\alpha$ and various coupling functions $h(\psi)$ on critical
temperature $T_c$ and condensation, can be obtained by solving
equations (\ref{eqpsi}) and (\ref{eqphi}) numerically by applying
shooting method. The results of critical temperatures are listed
in table 1. Based on these outcomes, increasing the effect of
non-minimal coupling in group 2(HCos, Exp(+) and P(+) cases) leads
to higher values of critical temperature while in group 1(F,
Exp(-) and P(-) cases) we face with difficulty in formation of
scalar hair. In addition, the trend of condensation as a function
of temperature for different strength of non-minimal coupling
constants in various coupling forms are shown in figure
\ref{fig1}. As said before, condensation values follow different
trends due to increasing the values of $\alpha$ for group 2 in
comparison with group 1. Conductor/superconductor phase transition
is easier to form for larger values of $\alpha$ in group 2 by
diminishing the values of condensation. Group 1 functions can not
give us information about low temperatures limit by increasing the
values of $\alpha$ because of numerical solution limitations.
However, we can follow numerical solution for $\alpha=15$ and even
larger values in P(+) case. Our choice of non-minimal coupling
function $h(\psi)$ is not so clear on the values of critical
temperature and we have two groups of answer which is reasonable
based on equations (\ref{ex1}) and (\ref{ex2}) for $\vert
\alpha\vert \psi^{2}\ll 1$ and we expect various results of
critical temperature for each case by increasing the values of
$\alpha$. However, running our numerical code for stronger effects
of $\alpha$ depends critically on our choice of $h(\psi)$. For
example, we can follow our numerical calculation just up to
$\alpha=0.4$ in group 1 while in group 2 it is so clear that each
forms of $h(\psi)$ allows us to obtain results up to a specific
value of non-minimal coupling constant. Furthermore, in all cases
we face second order phase transition.

\begin{table*}[tbp]
\begin{center}
\begin{tabular}{c|c|c|c|c|c|c|}
\cline{2-3}\cline{2-7}\cline{4-7}
& \multicolumn{3}{|c|}{group 1} & \multicolumn{3}{|c|}{group 2} \\
\cline{2-7}
& Exp(-) & F & P(-) & Exp(+) & HCos & P(+) \\
\hline
\multicolumn{1}{|c|}{$\alpha =10$} & $ --- $& $ --- $& $ --- $& $ --- $& $ --- $& $ 0.2415$ $\rho^{1/2}$ \\
\hline
\multicolumn{1}{|c|}{$\alpha =15$} & $ --- $& $ --- $& $ --- $& $ --- $& $ --- $& $0.2664 $ $\rho^{1/2}$ \\
\hline
\multicolumn{1}{|c|}{$\alpha =3$}& $ --- $& & $ --- $& $ --- $& $0.1828 $ $\rho^{1/2} $& $0.1828 $ $\rho^{1/2}$ \\
\hline
\multicolumn{1}{|c|}{$\alpha =2$}& $ --- $& $---$ & $ --- $& $ --- $& $0.1678 $ $\rho^{1/2} $& $0.1678 $ $\rho^{1/2}$ \\
\hline
\multicolumn{1}{|c|}{$\alpha =1$} & $ --- $ & $---$ & $ --- $& $ --- $& $ 0.1477 $ $\rho^{1/2}$ & $ 0.1477$ $\rho^{1/2}$ \\
\hline
\multicolumn{1}{|c|}{$\alpha =0.4$} & $ 0.1077$ $\rho^{1/2}$& $ 0.1077$ $\rho^{1/2}$ & $0.1077 $ $\rho^{1/2}$ & $0.1312 $ $\rho^{1/2}$ & $0.1312 $ $\rho^{1/2}$ & $0.1312 $ $\rho^{1/2}$ \\
\hline
\multicolumn{1}{|c|}{$\alpha =0.2$} & $0.1125 $ $\rho^{1/2}$ & $0.1125$ $\rho^{1/2}$ & $0.1125 $ $\rho^{1/2}$ & $0.1249 $ $\rho^{1/2}$ & $ 0.1249$ $\rho^{1/2}$ & $ 0.1249$ $\rho^{1/2}$\\
\hline
\multicolumn{1}{|c|}{$\alpha =0.0$} & $0.1184 $ $\rho^{1/2}$ & $0.1184 $ $\rho^{1/2}$ & $0.1184 $ $\rho^{1/2}$ & $0.1184 $ $\rho^{1/2}$ & $0.1184 $ $\rho^{1/2}$ & $ 0.1184$ $\rho^{1/2}$ \\
\hline
\end{tabular}%
\caption{Numerical results for critical temperature $T_c$ in $d=4$ for different values of $\alpha$ in each form of $h(\psi)$.}
\end{center}
\end{table*}\label{tab1}

\subsection*{Conductivity}
\begin{figure*}[tbp]
\centering
\subfigure[~Re]{\includegraphics[width=0.4\textwidth]{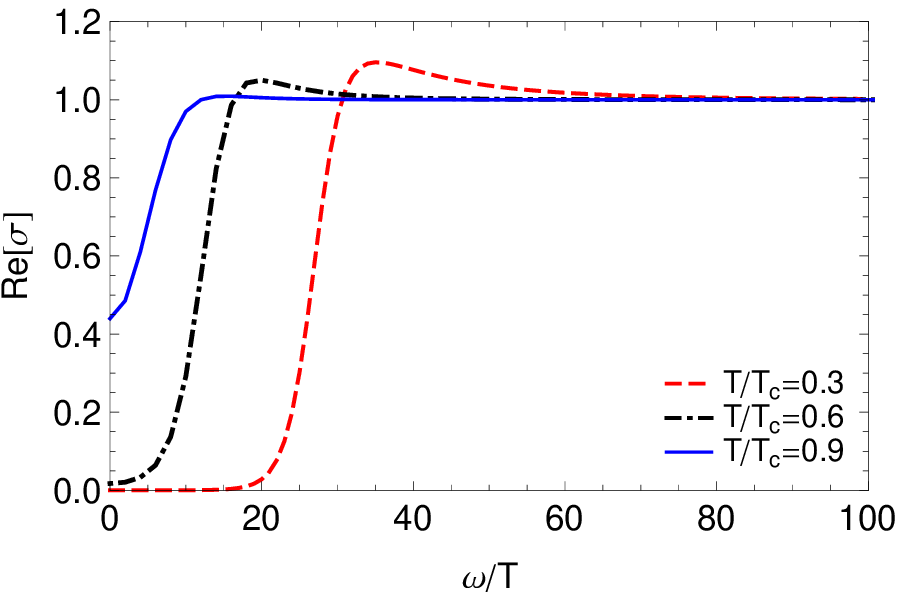}} \qquad %
\subfigure[~Im]{\includegraphics[width=0.4\textwidth]{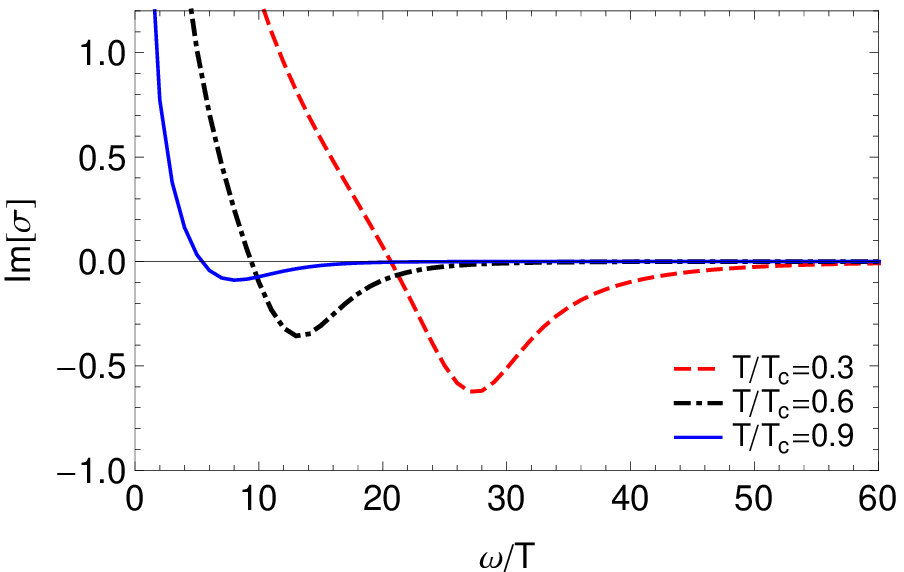}} \qquad %

\caption{The behavior of conductivity as a function of temperature with $\alpha=0$.}\label{fig2}
\end{figure*}

\begin{figure*}[tbp]
\centering
\subfigure[~Exp(-)]{\includegraphics[width=0.4\textwidth]{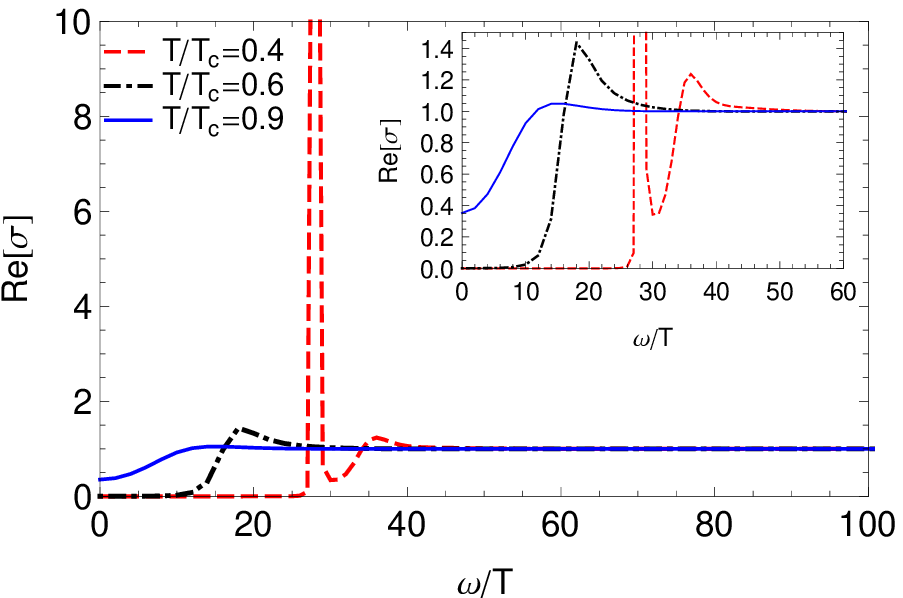}} \qquad %
\subfigure[~Exp(+)]{\includegraphics[width=0.4\textwidth]{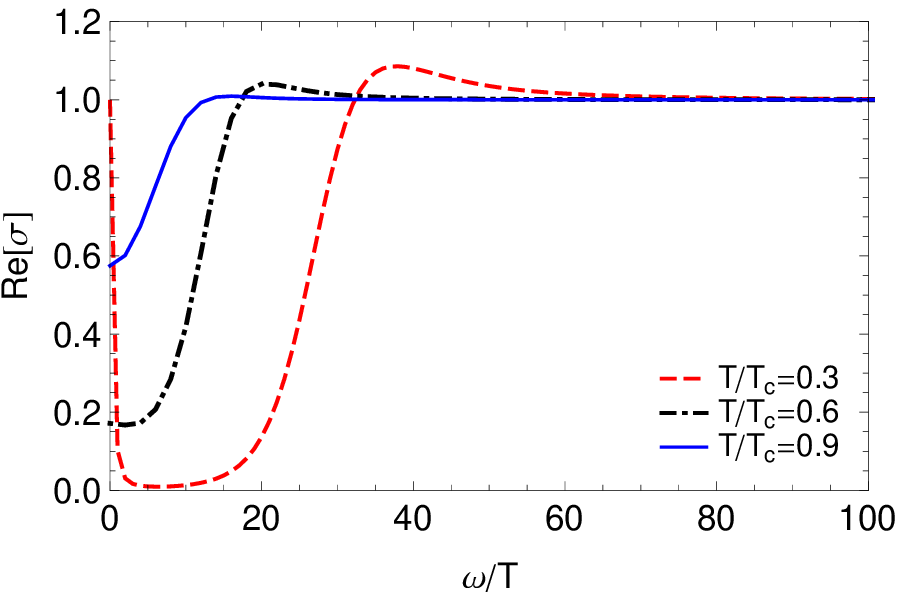}} \qquad %
\subfigure[~F]{\includegraphics[width=0.4\textwidth]{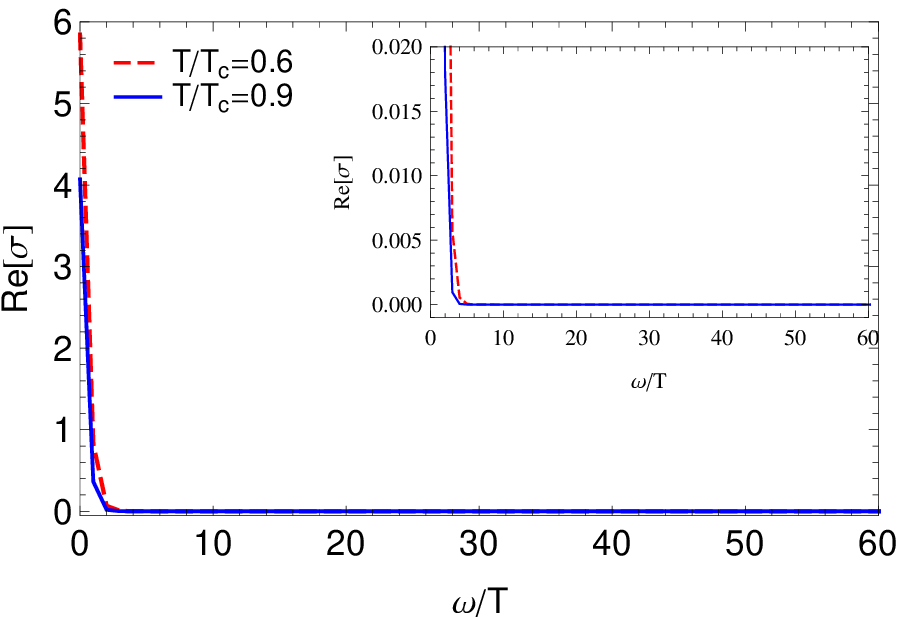}} \qquad %
\subfigure[~HCos]{\includegraphics[width=0.4\textwidth]{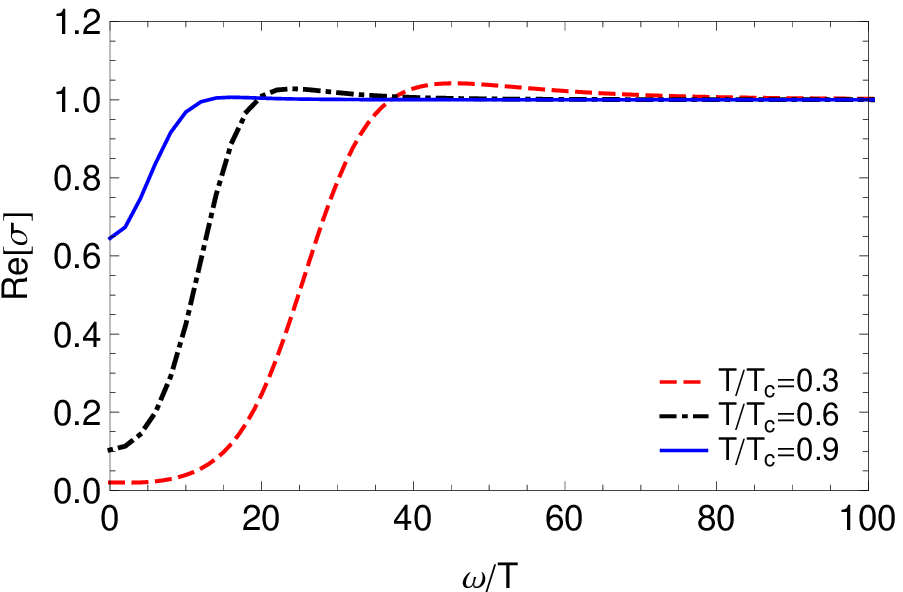}} \qquad %
\subfigure[~P(-)]{\includegraphics[width=0.4\textwidth]{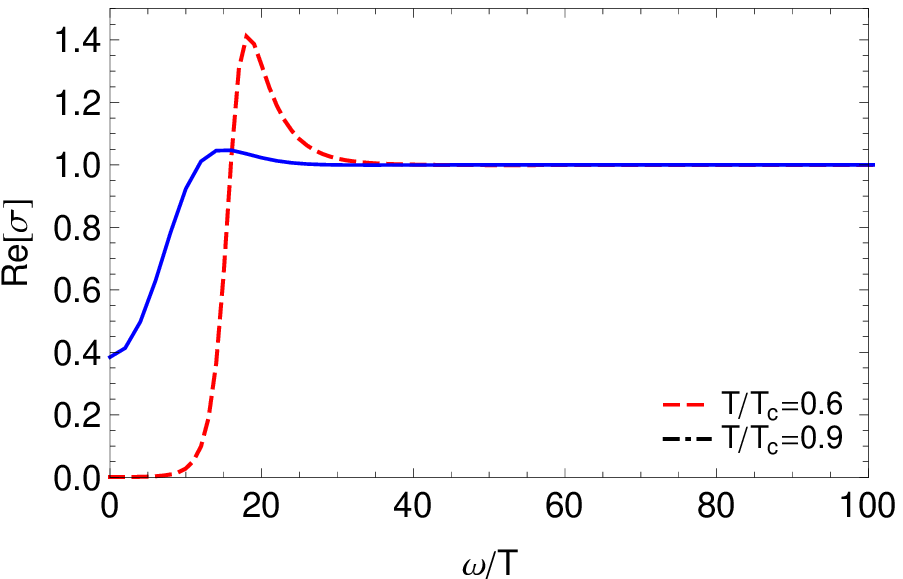}} \qquad %
\subfigure[~P(+)]{\includegraphics[width=0.4\textwidth]{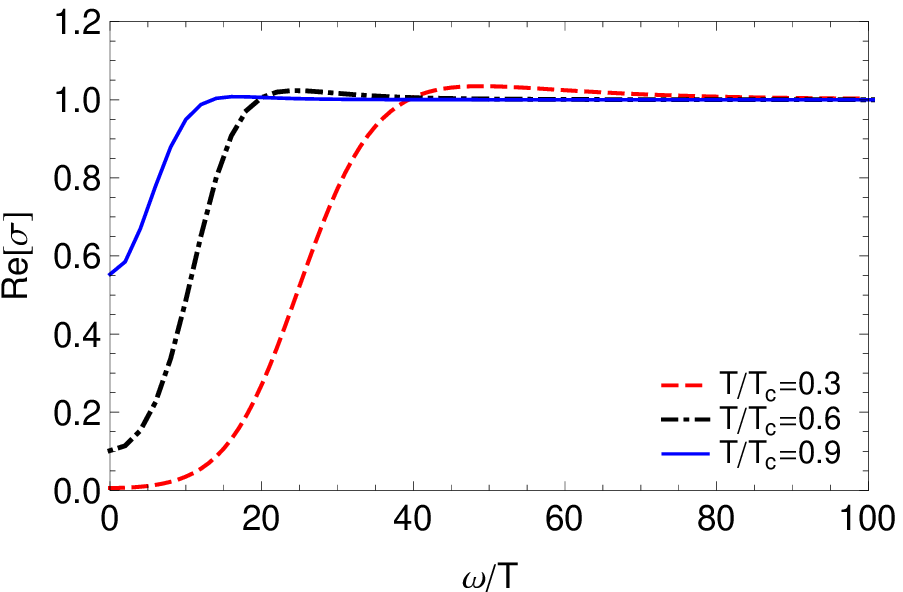}} \qquad %

\caption{The behavior of the real part of conductivity as a function of temperature with $\alpha=0.4$.}\label{fig3}
\end{figure*}

\begin{figure*}[tbp]
\centering
\subfigure[~Exp(-)]{\includegraphics[width=0.4\textwidth]{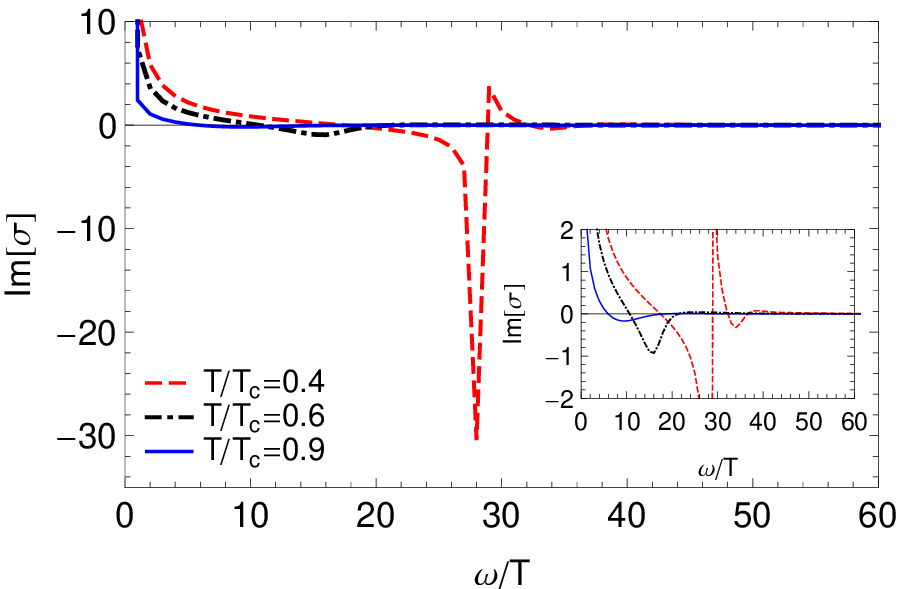}} \qquad %
\subfigure[~Exp(+)]{\includegraphics[width=0.4\textwidth]{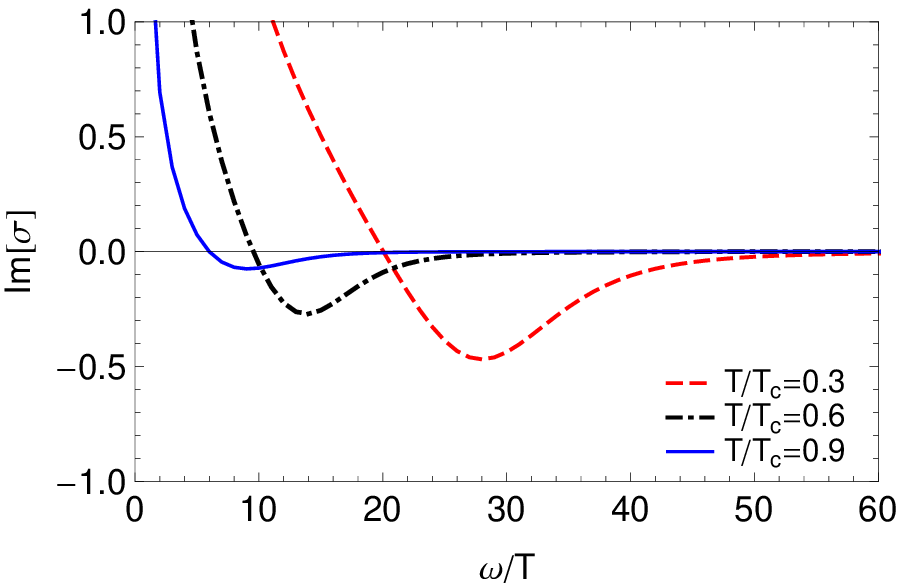}} \qquad %
\subfigure[~F]{\includegraphics[width=0.4\textwidth]{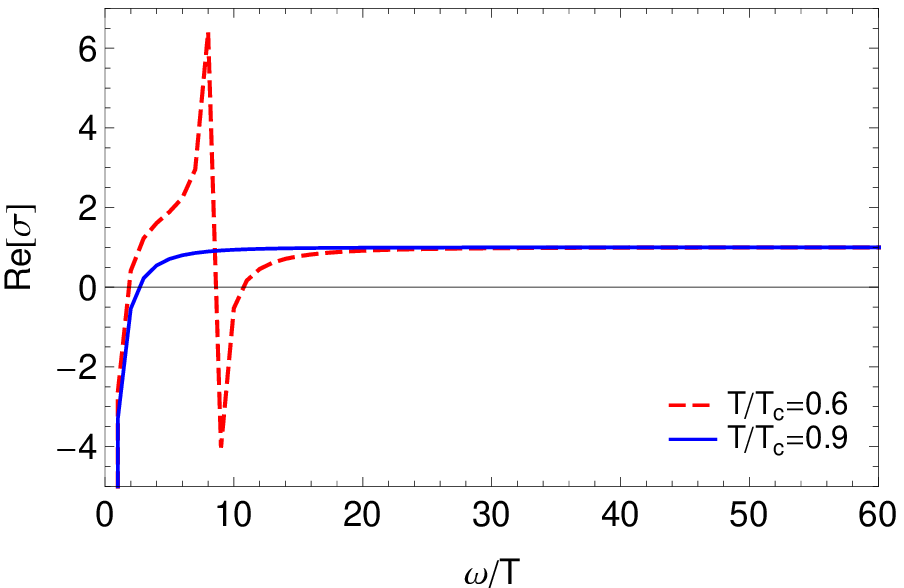}} \qquad %
\subfigure[~HCos]{\includegraphics[width=0.4\textwidth]{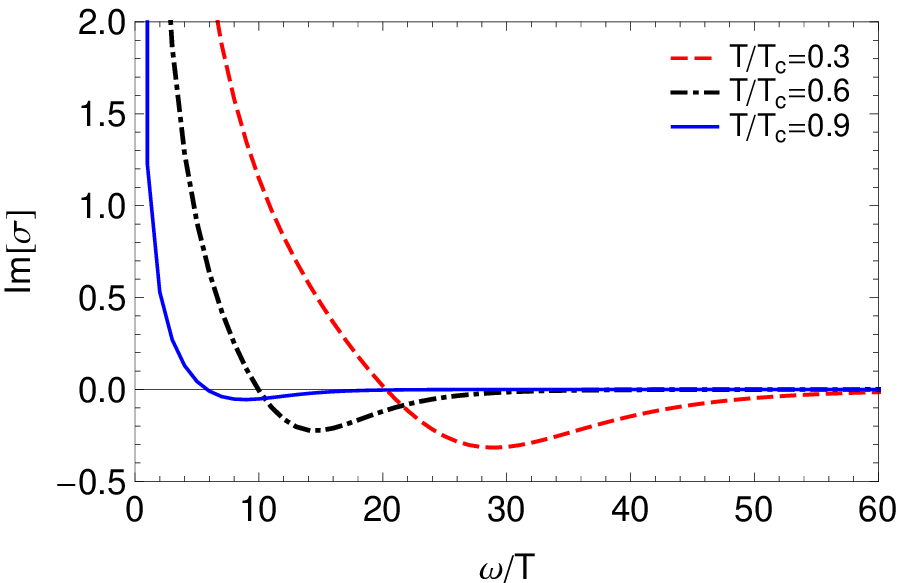}} \qquad %
\subfigure[~P(-)]{\includegraphics[width=0.4\textwidth]{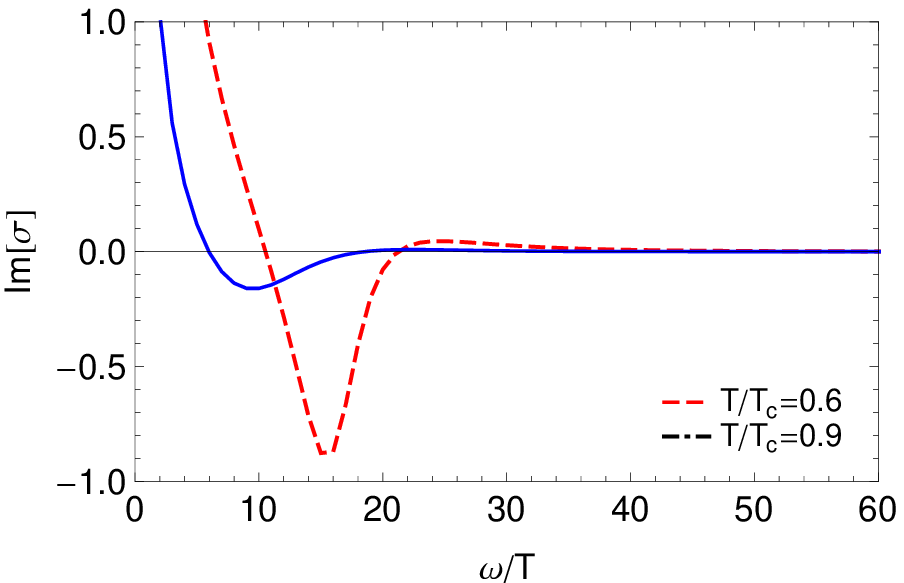}} \qquad %
\subfigure[~P(+)]{\includegraphics[width=0.4\textwidth]{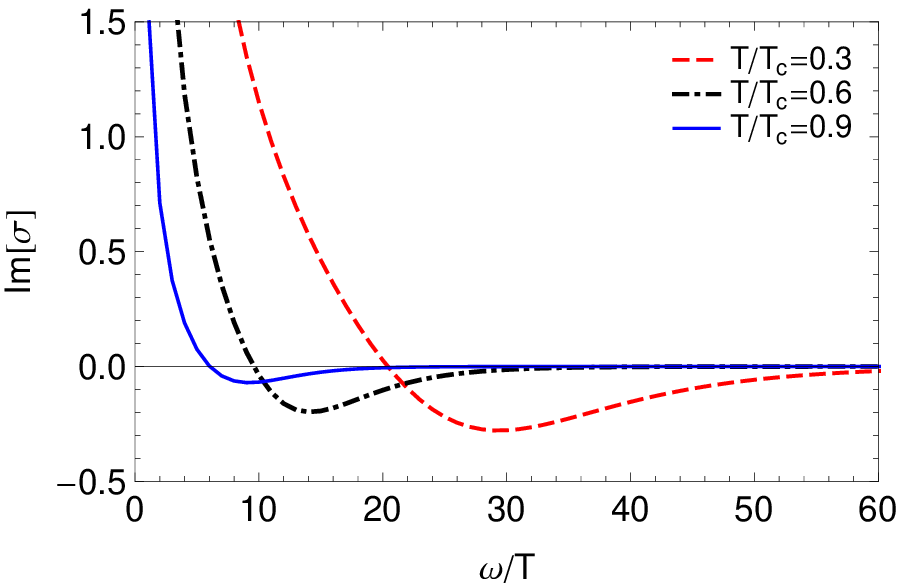}} \qquad %

\caption{The behavior of the imaginary part of conductivity as a function of temperature with $\alpha=0.4$.}\label{fig4}
\end{figure*}

\begin{figure*}[tbp]
\centering
\subfigure[~Exp(-), $T=0.4T_C$]{\includegraphics[width=0.4\textwidth]{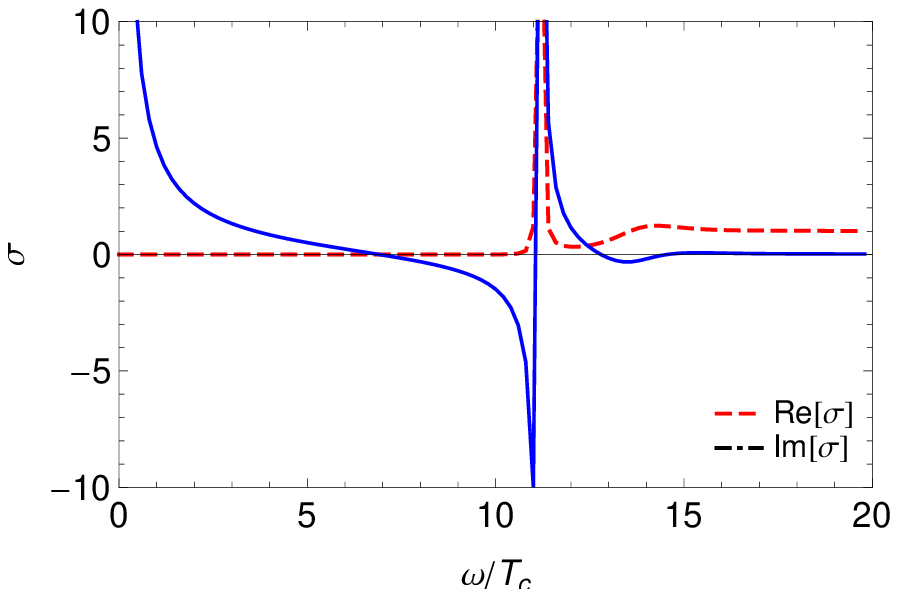}} \qquad %
\subfigure[~HCos, $T=0.3T_C$]{\includegraphics[width=0.4\textwidth]{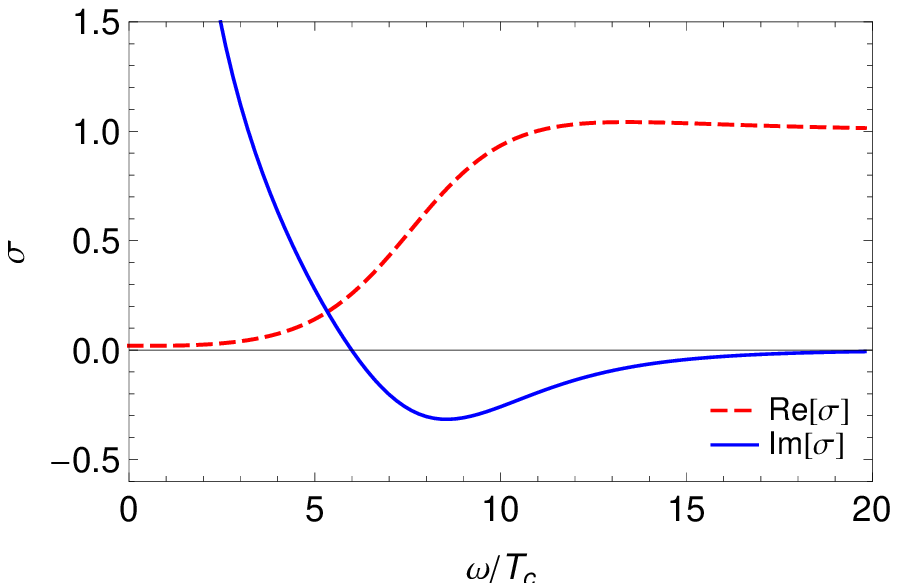}} \qquad %
\subfigure[~P(+), $T=0.3T_C$]{\includegraphics[width=0.4\textwidth]{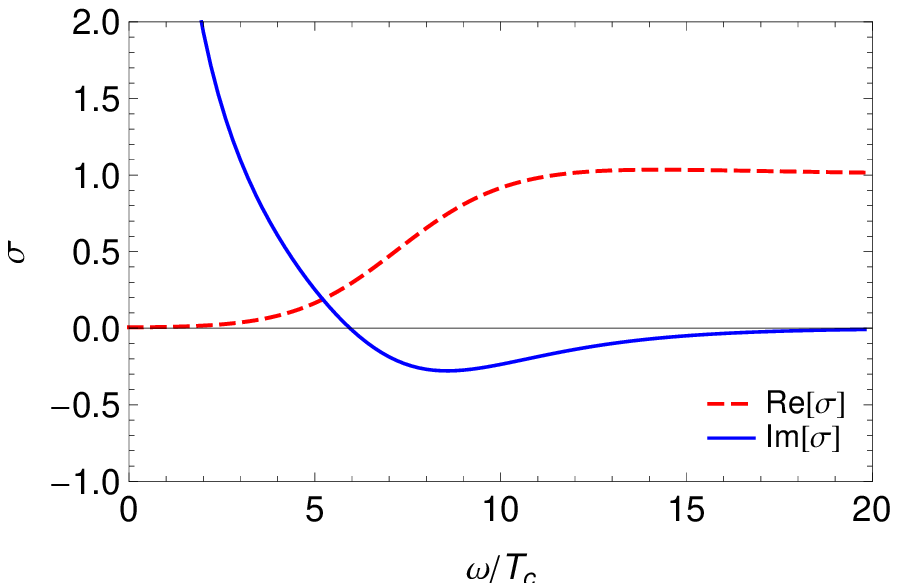}} \qquad %

\caption{The behavior of conductivity as a function of temperature with $\alpha=0.4$.}\label{fig5}
\end{figure*}

\begin{figure*}[tbp]
\centering
\subfigure[~HCos]{\includegraphics[width=0.4\textwidth]{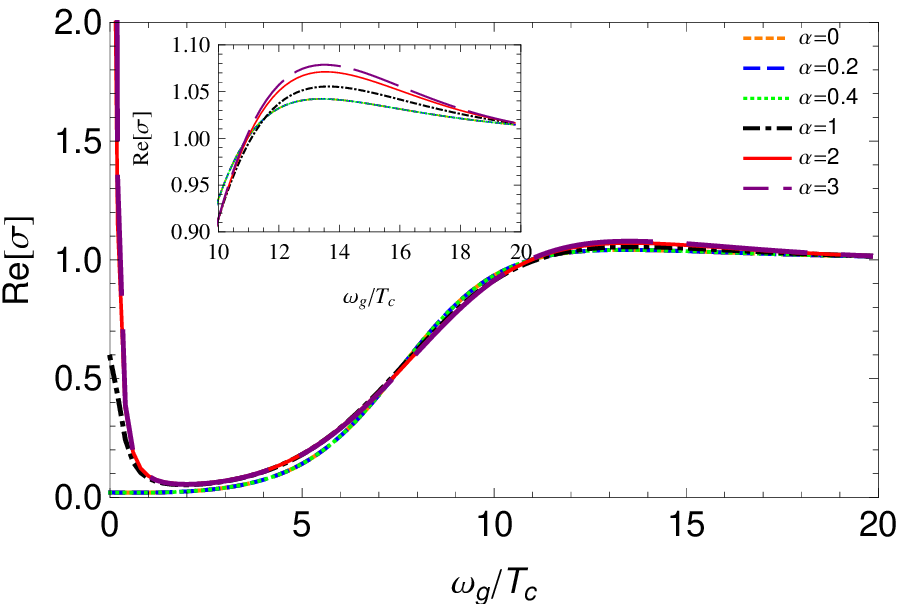}} \qquad %
\subfigure[~HCos]{\includegraphics[width=0.4\textwidth]{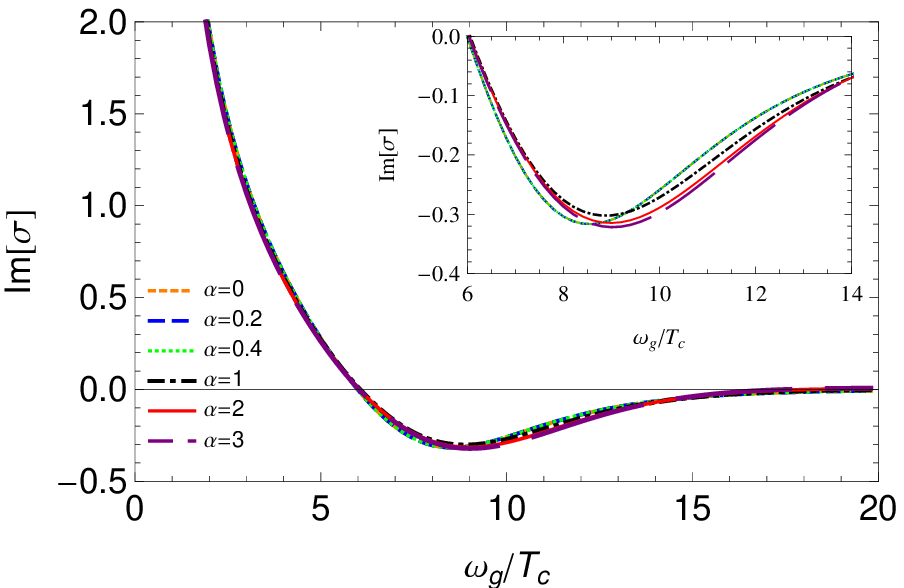}} \qquad %
\subfigure[~P(+)]{\includegraphics[width=0.4\textwidth]{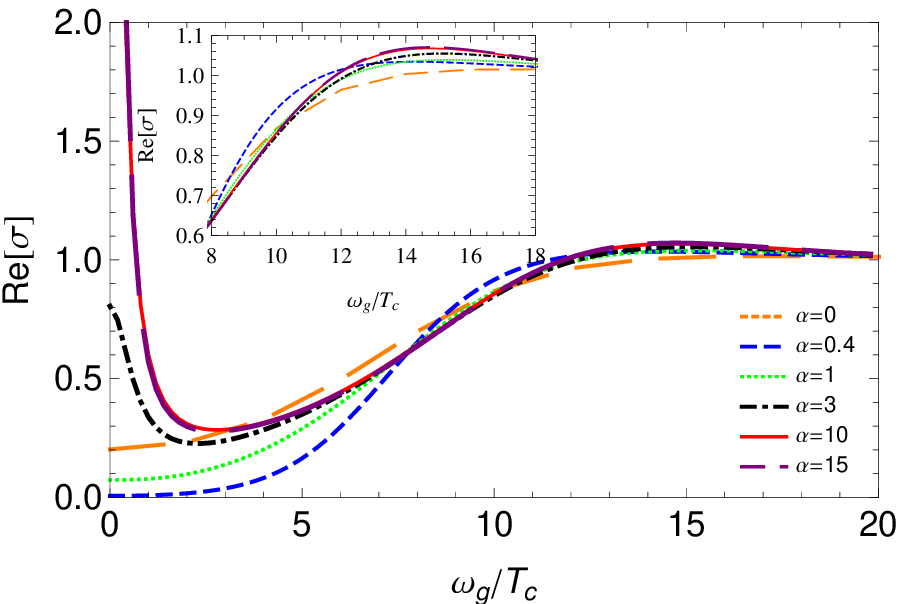}} \qquad %
\subfigure[~P(+)]{\includegraphics[width=0.4\textwidth]{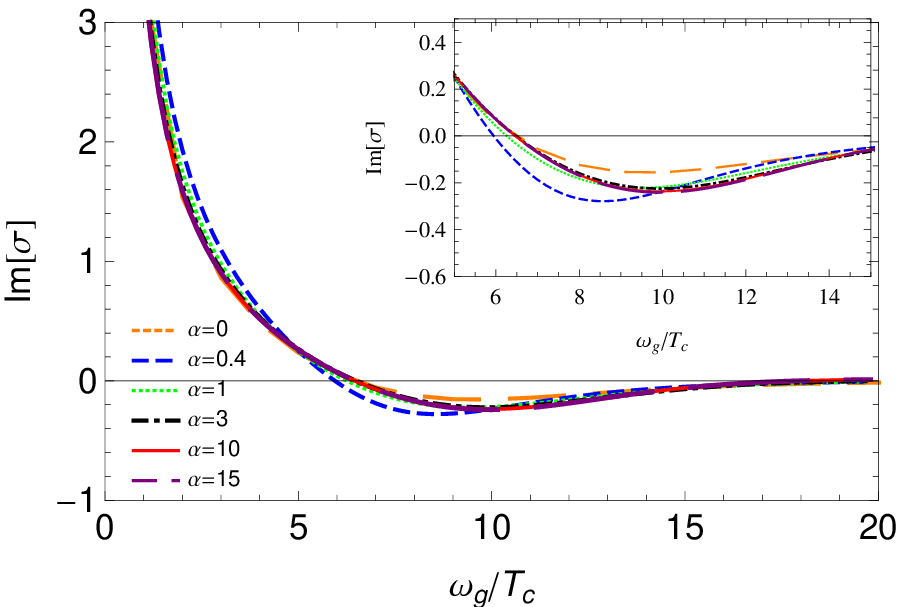}} \qquad %

\caption{The behavior of conductivity parts as a function of temperature with $T=0.3T_C$.}\label{fig6}
\end{figure*}

Electrical conductivity in holographic setup can be obtained by
applying an appropriate perturbation as $\delta A_x = A_x e^{- i
\omega t}$ on the black hole background which yields to
\begin{equation}
A_{x}''+ \left[\frac{f'}{f}+\frac{h'(\psi)}{h(\psi)}\right]A_{x}'+ \left[\frac{\omega ^2}{f^2}-\frac{2 q^2 \psi^2 }{f h(\psi)}\right]A_{x}=0,
\end{equation}
and behaves asymptotically as
\begin{equation}
A_{x}''+\frac{2}{r}A_{x}'+\frac{\omega ^2 }{r^4}A_{x}=0,
\end{equation}
by considering $A^{(0)}$ and $A^{(1)}$ as constant parameters, asymptotic solution of $A_x$ is
\begin{equation} \label{aysolgauss}
A_{x} =A^{(0)}+\frac{A^{(1)}}{r}+\cdots,
\end{equation}%
so, the electrical conductivity can be evaluated as
\begin{equation} \label{conductivitygauss}
\sigma_{xx} =\frac{ A^{(1)}}{i \omega A^{(0)}}.
\end{equation}
Behavior of real and imaginary parts of conductivity as a function
of $\omega/T$ is shown in figures \ref{fig2}-\ref{fig6} for
different values of nonminimal coupling in various forms of
$h(\psi)$. In general figures follow some universal trends. For
example at low frequency regime, real and imaginary parts of
conductivity are related to each other via Kramers-Kronig relation
by having a delta and pole respectively. However, at high
frequency limit, both parts of conductivity tend to a constant
value. We observe the appearance of gap frequency at low
temperatures which shifts to higher values of $\omega_g/T_c$ by
increasing the strength of non-minimal coupling from $8$ to $11$
which is much larger than the BCS prediction $\omega_g\sim 3.5T_c$
due to strong coupling nature of high temperature superconductors.
Because of difficulty in numerical solution of P(-) and F cases,
we can not obtain their behavior of conductivity at $T=0.3T_c$.
Meanwhile, the trend of conductivity in F case is far different
from other models.


\section{Holographic setup with hyperscaling violation}\label{sec2.2} 
In order to study the effect of hyperscaling violation on the
holographic superconductors, the black hole solution with
hyperscaling violation reads \cite{hyper1,hyper5}

\begin{equation}
ds^{2}=r^{\frac{-2 \theta}{d}}\left(-r^{2z} f(r) dt^{2}+\frac{1}{r^2 f(r)}dr^2+r^{2}(dx^2+dy^2) \right),
\end{equation}
where
\begin{equation}
f(r)={\left( 1-\frac{r_+}{r} \right)}^{2+z-\theta},
\end{equation}
with $r_+$, $z$ and $\theta$ as the radius of the event horizon, dynamical exponent and hyperscaling
violation exponent, respectively. Therefore, the Hawking temperature is given by \cite{hyper5}
\begin{equation}
T=\frac{f'(r_+)}{4 \pi}=\frac{2+z-\theta}{4 \pi}.
\end{equation}
We concentrate on probe limit because it can simplify the problem but retain most of the interesting physics.
At the asymptotic limit $r\rightarrow \infty$ and $f(r)\rightarrow 1$, the metric reduces to \cite{hyper5}
\begin{equation}
ds^{2}=r^{\frac{-2 \theta}{d}}\left(-r^{2z} dt^{2}+\frac{1}{r^2}dr^2+r^{2}(dx^2+dy^2) \right),
\end{equation}
this is the most general geometry which is spatially homogeneous and covariant under the following scale transformations \cite{hyper5}
\begin{equation}
t\rightarrow \lambda^z t, \ \ \ r\rightarrow \lambda^{-1} r, \ \ \ x_i\rightarrow \lambda x_i, \ \ \ ds_{d+2} \rightarrow \lambda^{\frac{\theta}{d}} ds_{d+2},
\end{equation}
in which it is obvious that for non-zero values of $\theta$, the distance is not invariant under the scaling which in the context
of AdS/CFT indicates violations of hyperscaling in the dual field theory \cite{hyper5}.
Evidently when $\theta=0$ and $z\neq 1$, it reduces to the pure Lifshitz case \cite{LU}.
In this work, we focus on $d = 2$, $z = 2$.
Observations indicate that backgrounds whose asymptotic behavior coincides with the
above metric may be of interest to condensed matter physics. Therefore, studying AdS/CFT correspondence in this case is valuable.

\section*{Condensation of scalar field}
\begin{figure*}[tbp]
\centering
\subfigure[~$\alpha=0$]{\includegraphics[width=0.4\textwidth]{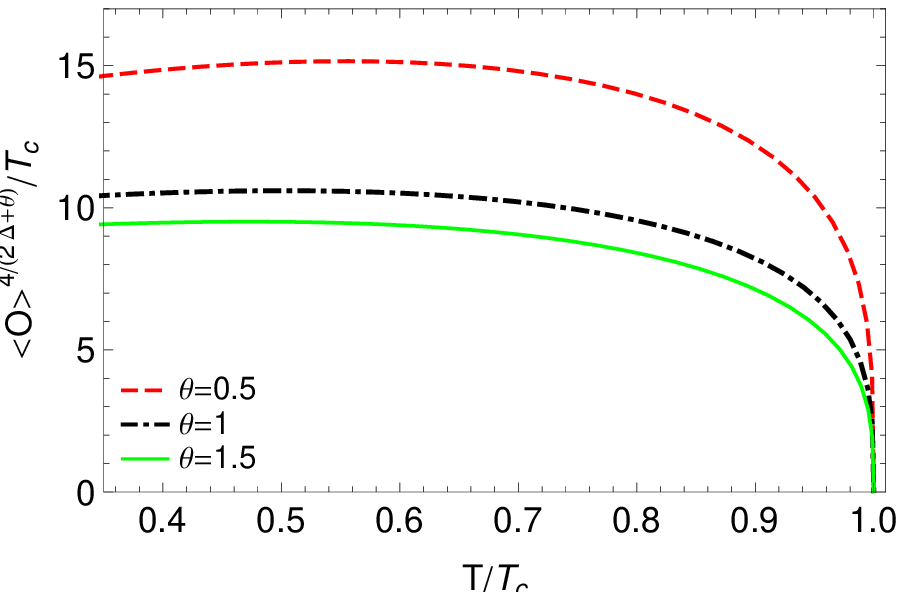}} \qquad %
\subfigure[~$\alpha=3$]{\includegraphics[width=0.4\textwidth]{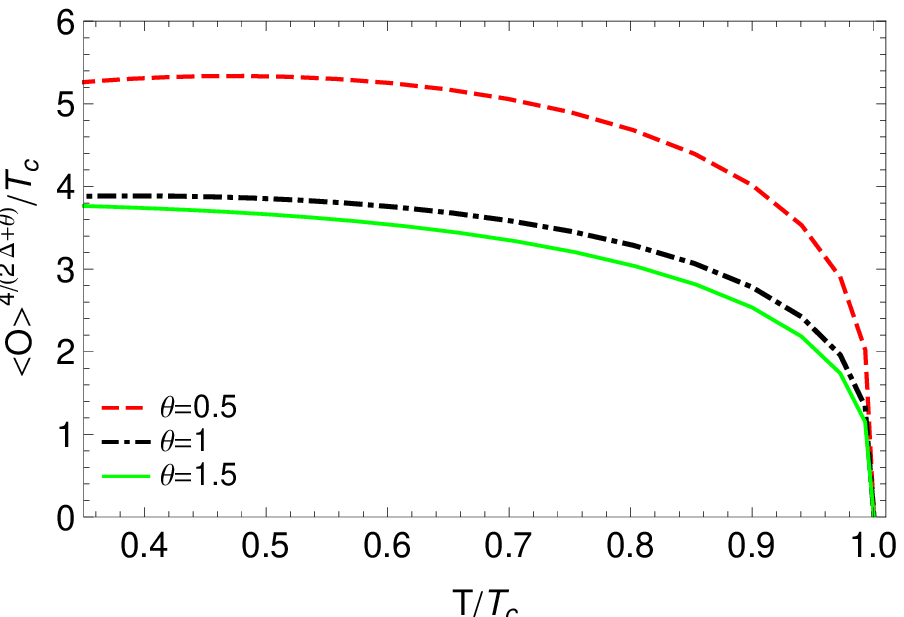}} \qquad %
\subfigure[~$\theta=0$]{\includegraphics[width=0.4\textwidth]{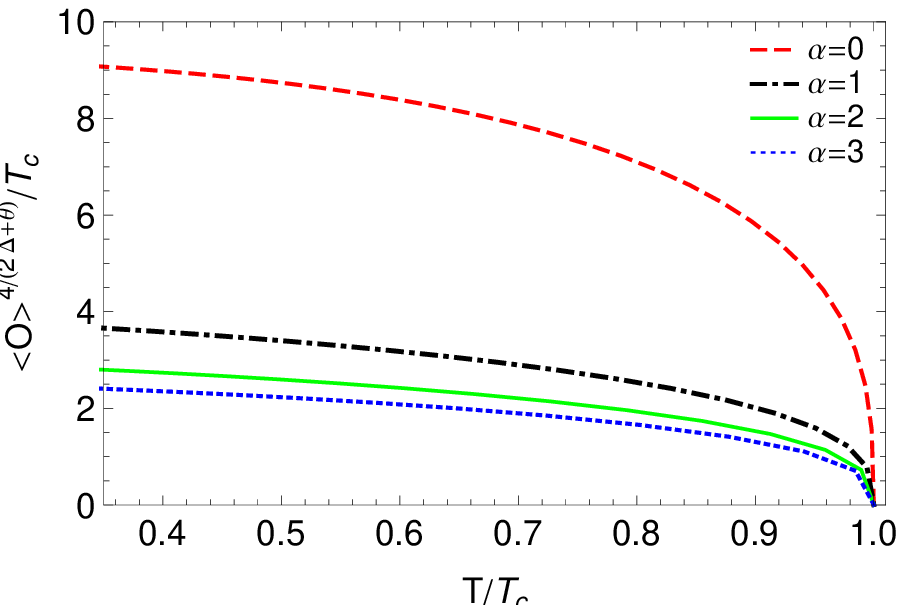}} \qquad %
\subfigure[~$\theta=1.5$]{\includegraphics[width=0.4\textwidth]{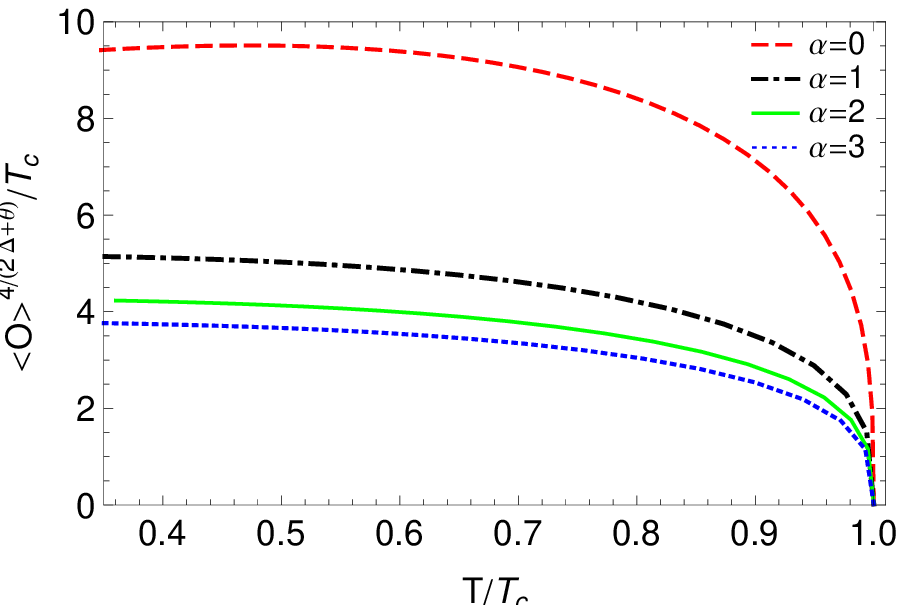}} \qquad %
\caption{The behavior of the condensation as a function of
temperature.}\label{fig1.1}
\end{figure*}

In the probe limit, we will construct the holographic
superconductors when a real scalar field coupled non-minimally to
the electromagnetic gauge field in the background of the
hyperscaling violation black hole as equation (\ref{eqact})
\cite{05,06,07} with $h(\psi)=1+\alpha \psi^2$. We obtain the
equations of motion by varying the action with respect to gauge
field $A_{\mu}$ and scalar field $\psi$ and taking the ansatz of
the matter fields as $A_{\mu}=\phi(r)$ and $\psi=\psi(r)$
\begin{equation}\label{eqphii}
\phi ''+\left[\frac{2 \alpha \psi \psi '}{1+\alpha \psi^2}-\frac{z}{r}+\frac{3}{r}\right] \phi '-\frac{2 q^2 \psi ^2 }{f \left(1+\alpha \psi ^2\right)r^{\theta+2}}\phi =0,
\end{equation}
\begin{equation}\label{eqpsii}
\psi ''+\left[-\frac{\theta}{r}+\frac{f'}{f}+\frac{z}{r}+\frac{3}{r}\right]\psi '+\left[-\frac{m^2 }{f r^{\theta+2}}+\frac{\alpha \phi '^2}{2 f r^{\theta +2 z}}+\frac{q^2 \phi^2}{f^2 r^{2 z+2}}\right] \psi =0,
\end{equation}
which have asymptotic behavior as \cite{hyper4}
\begin{equation}
\phi=\rho+\mu \ ln(r),
\end{equation}
\begin{equation} \label{eqasy}
\psi(r) = \left\{
\begin{array}{lr}
\psi_- r^{4-\Delta} +\psi_+ r^{\Delta}, \ \ \ \Delta=2+\sqrt{4+m^2} \ \ \ \text{with} \ \ \ \theta=0, \\
\bigskip\\
\psi_- +\psi_+ r^{\Delta}, \ \ \ \Delta=4-\theta \ \ \ \text{with} \ \ \ 0<\theta<2, \\
\end{array} \right.
\end{equation}%
Same as previous section by applying shooting method, we study the effect of different values of hyperscaling term $\theta$ and non-minimal coupling constant $\alpha$ by solving equations (\ref{eqphii}) and (\ref{eqpsii}) numerically. We choose $m^2=-3$. In table II, the consequences of different choices of $\theta$ and $\alpha$ in critical temperature $T_c$ are listed. Based on our results, increasing each of these parameters makes conductor/superconductor phase transition easier to form by increasing the critical temperature. In addition, the effects of the hyperscaling violation on the condensation of the scalar operators is shown in figure \ref{fig1.1} which prove the outcomes of critical temperature. In all cases for higher effects of hyperscaling and non-minimal terms, condensation values go down. Moreover, for all cases considered here, the condensation is single-valued near the critical temperature and the condensate drops to zero continuously at the critical
temperature. Fitting these curves, we obtain a square root behavior for small condensate $<O>=\left(1-T/T_c\right)^{1/2}
$ which means that the phase transition belongs to the second order and the critical exponent of the system
takes the mean-field value 1/2.
\begin{table*}[tbp]
\label{tab5}
\begin{center}
\begin{tabular}{c|c|c|c|c|}
\cline{2-5}
& $\alpha=0$ &$\alpha=1$ & $\alpha=2$ & $\alpha=3$ \\
\hline
\multicolumn{1}{|c|}{$\theta=0$} & $0.0352 $ $\mu$& $0.0809 $ $\mu$ & $ 0.1103 $ $\mu$ & $0.1333 $ $\mu$ \\
\hline
\multicolumn{1}{|c|}{$\theta=0.5$} & $0.0331 $ $\mu$ & $ 0.0825$ $\mu$ & $ 0.1127$ $\mu$ & $ 0.1364$ $\mu$ \\
\hline
\multicolumn{1}{|c|}{$\theta=1$} & $ 0.0335 $ $\mu$ & $0.0870 $ $\mu$ & $ 0.1190$ $\mu$ & $ 0.1440$ $\mu$ \\
\hline
\multicolumn{1}{|c|}{$\theta=1.5$} & $ 0.0361 $ $\mu$ & $0.0963 $ $\mu$ & $ 0.1316$ $\mu$ & $0.1593 $ $\mu$ \\
\hline
\end{tabular}%
\caption{Numerical results of critical temperature for different
values of hyperscaling parameters $\theta$ and non-minimal coupling term
$\alpha$.}
\end{center}
\end{table*}
Except when $\alpha=0$, the critical temperature first goes down
and then goes up by increasing $\theta$.
\section*{Conductivity}
\begin{figure*}[tbp]
\centering
\subfigure[~$\alpha=0$, $\theta=0$]{\includegraphics[width=0.4\textwidth]{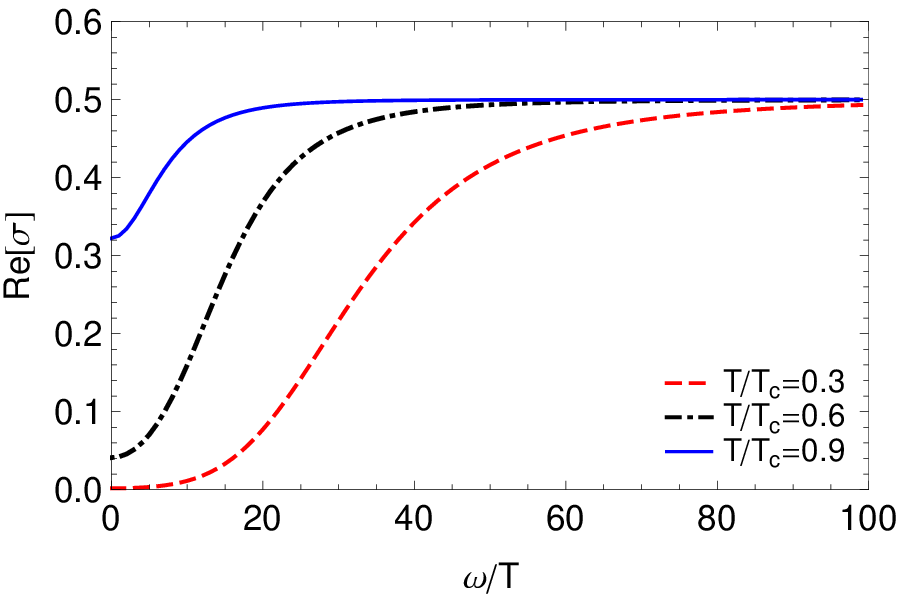}} \qquad %
\subfigure[~$\alpha=3$, $\theta=0$]{\includegraphics[width=0.4\textwidth]{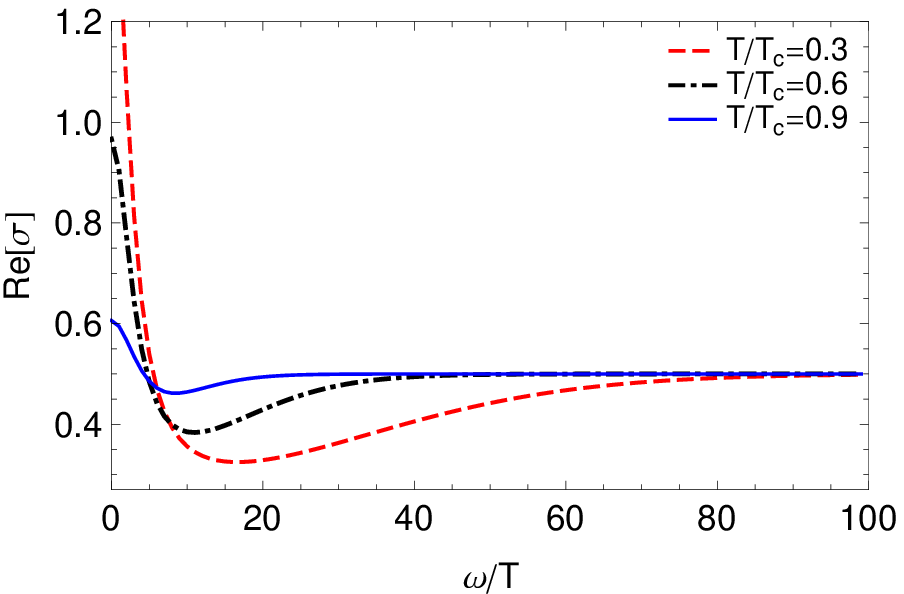}} \qquad %
\subfigure[~$\alpha=0$, $\theta=1.5$]{\includegraphics[width=0.4\textwidth]{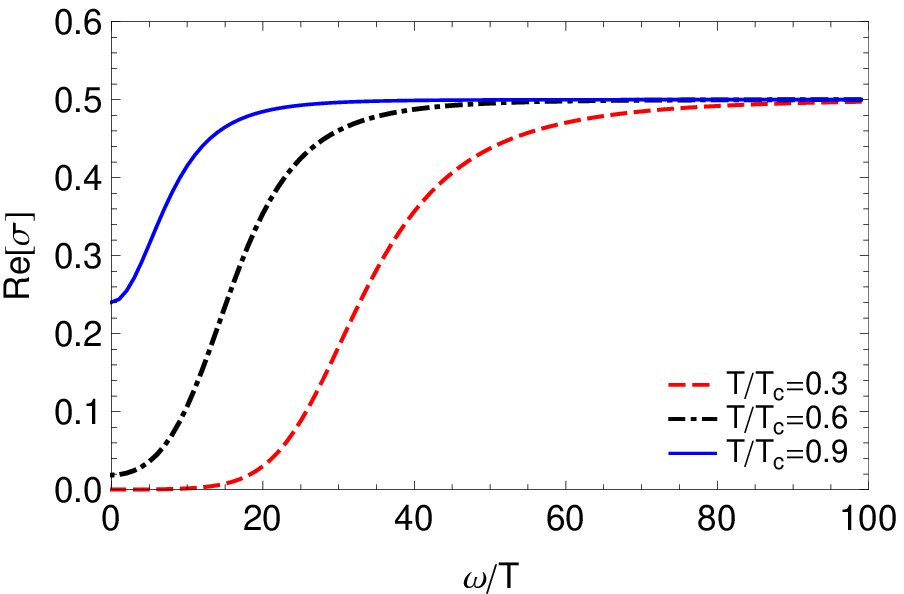}} \qquad %
\subfigure[~$\alpha=3$, $\theta=1.5$]{\includegraphics[width=0.4\textwidth]{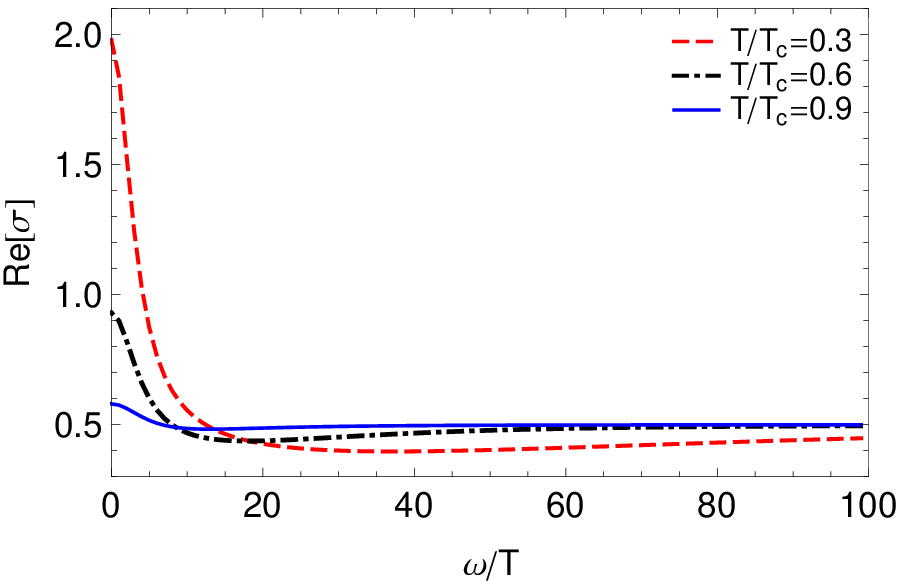}} \qquad %
\caption{The behavior of real part of conductivity.}\label{fig2.1}
\end{figure*}

\begin{figure*}[tbp]
\centering
\subfigure[~$\alpha=0$, $\theta=0$]{\includegraphics[width=0.4\textwidth]{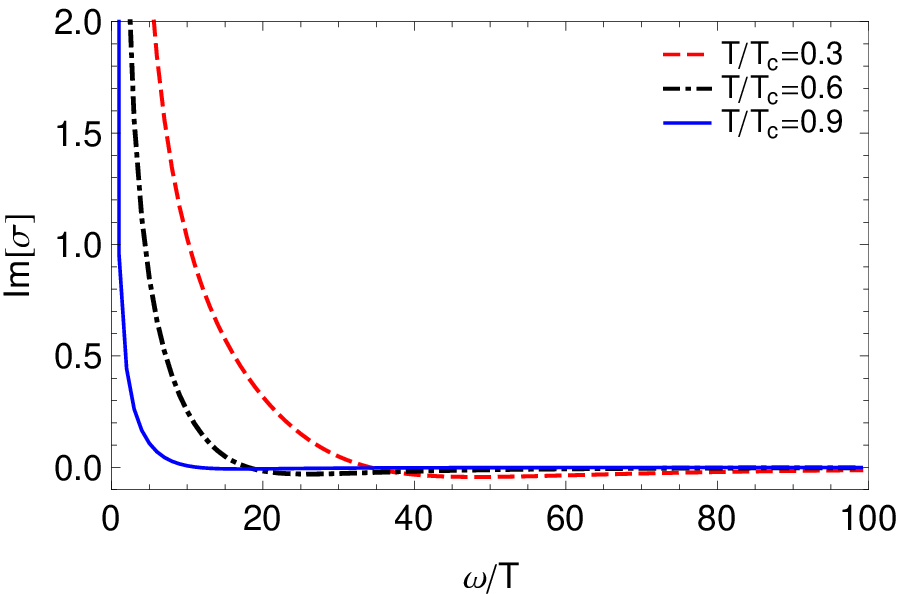}} \qquad %
\subfigure[~$\alpha=3$, $\theta=0$]{\includegraphics[width=0.4\textwidth]{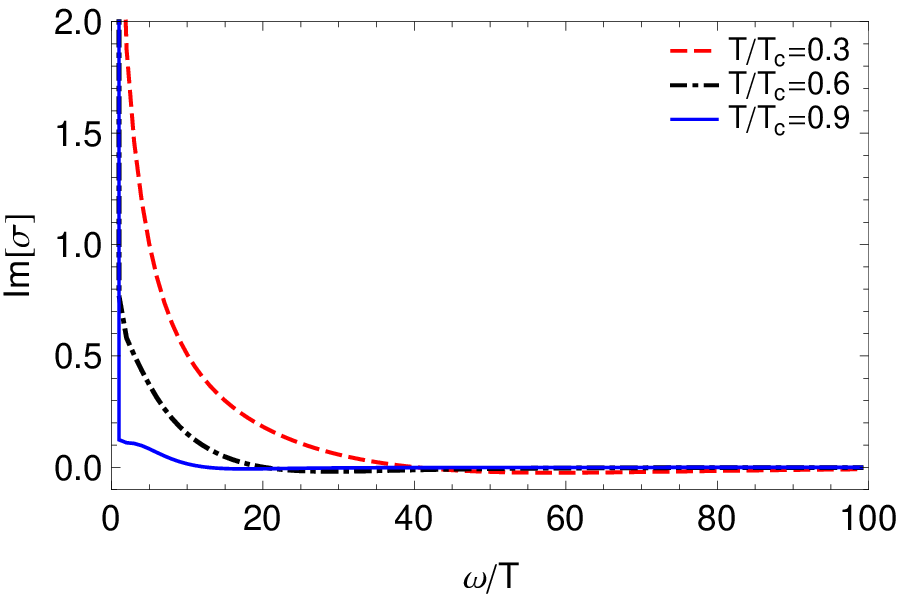}} \qquad %
\subfigure[~$\alpha=0$, $\theta=1.5$]{\includegraphics[width=0.4\textwidth]{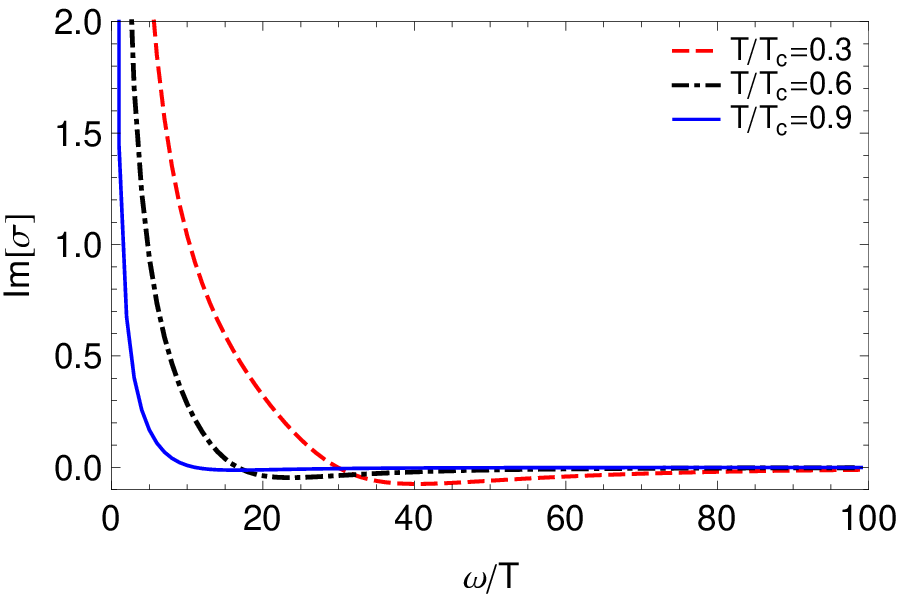}} \qquad %
\subfigure[~$\alpha=3$, $\theta=1.5$]{\includegraphics[width=0.4\textwidth]{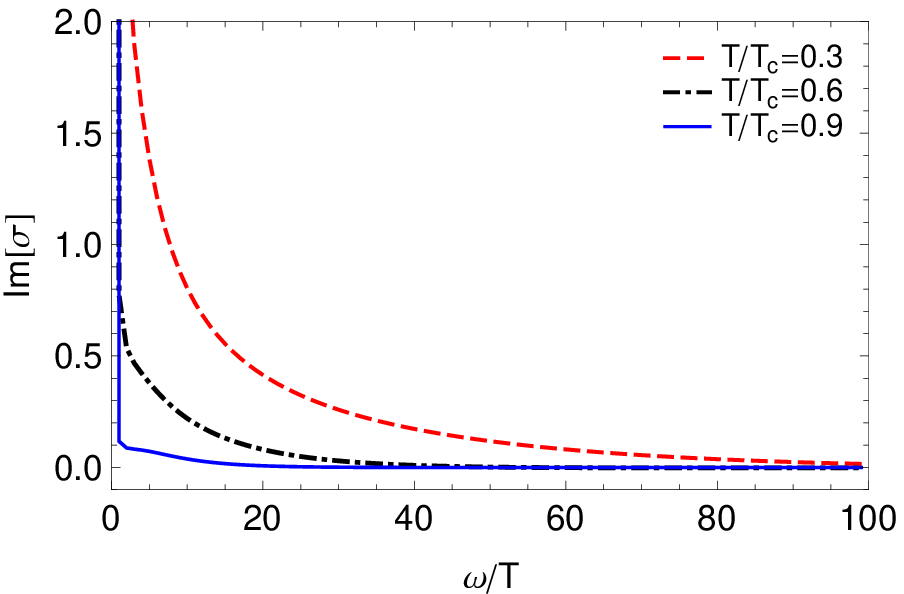}} \qquad %
\caption{The behavior of imaginary part of conductivity.}\label{fig3.1}
\end{figure*}

\begin{figure*}[tbp]
\centering
\subfigure[~$\alpha=0$]{\includegraphics[width=0.4\textwidth]{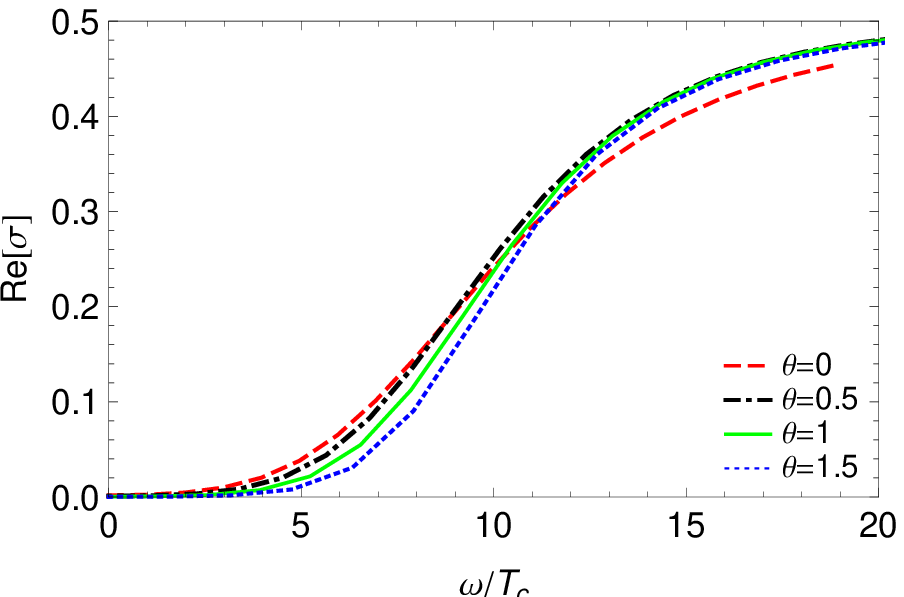}} \qquad %
\subfigure[~$\alpha=3$]{\includegraphics[width=0.4\textwidth]{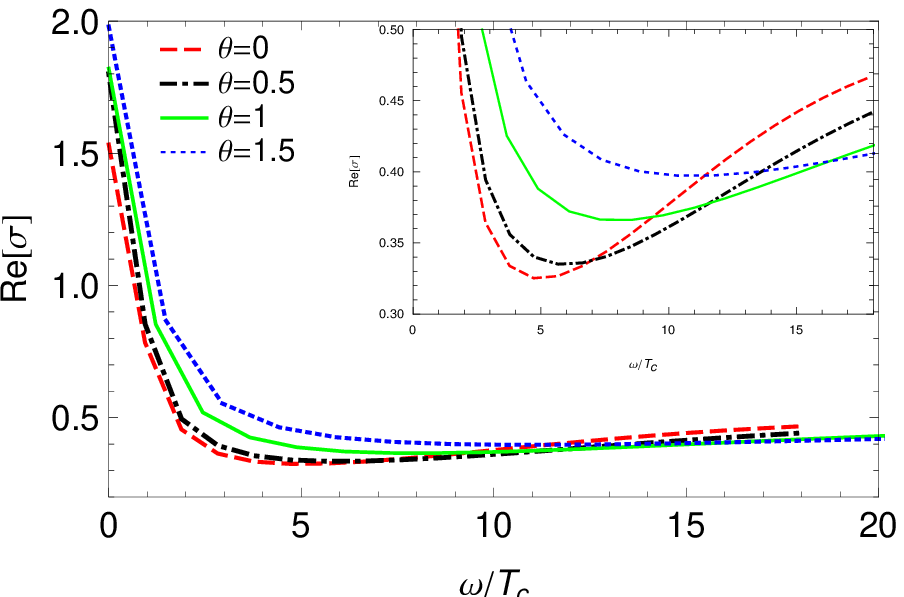}} \qquad %
\subfigure[~$\theta=0$]{\includegraphics[width=0.4\textwidth]{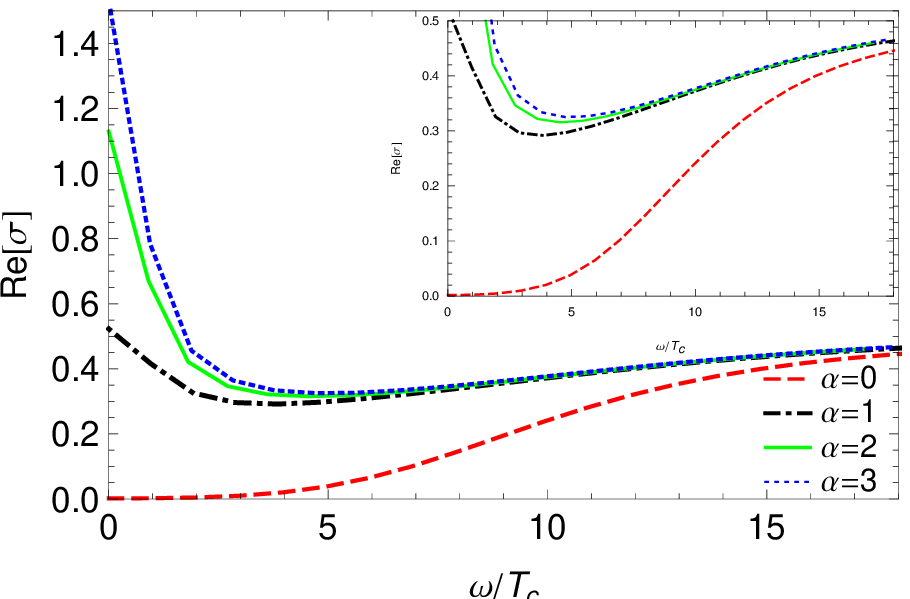}} \qquad %
\subfigure[~$\theta=1.5$]{\includegraphics[width=0.4\textwidth]{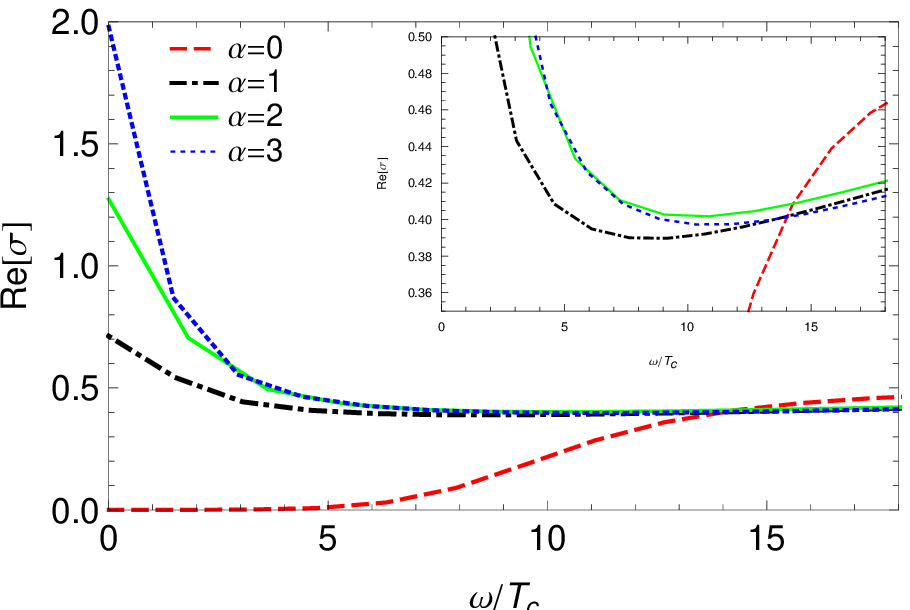}} \qquad %
\caption{The behavior of real part of conductivity in $T=0.3T_c$.}\label{fig4.1}
\end{figure*}

\begin{figure*}[tbp]
\centering
\subfigure[~$\alpha=0$]{\includegraphics[width=0.4\textwidth]{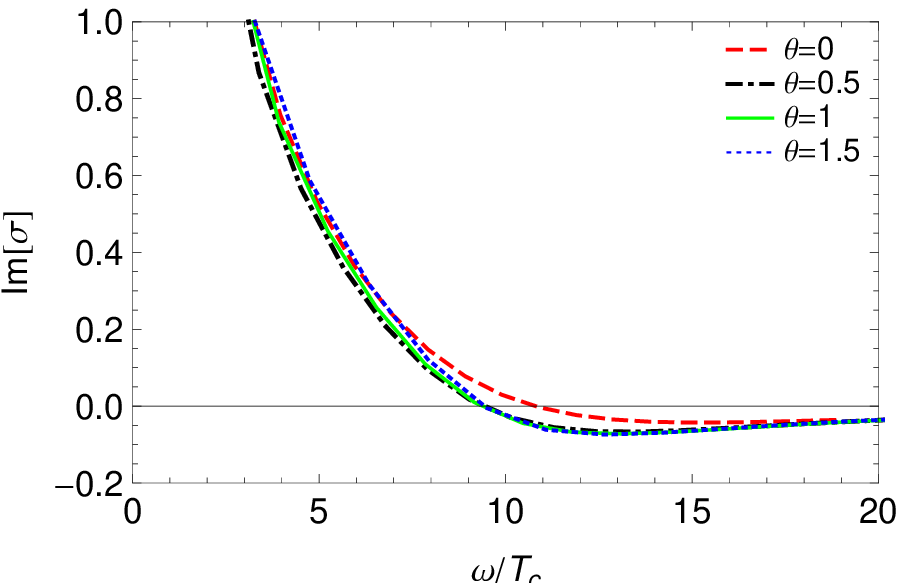}} \qquad %
\subfigure[~$\alpha=3$]{\includegraphics[width=0.4\textwidth]{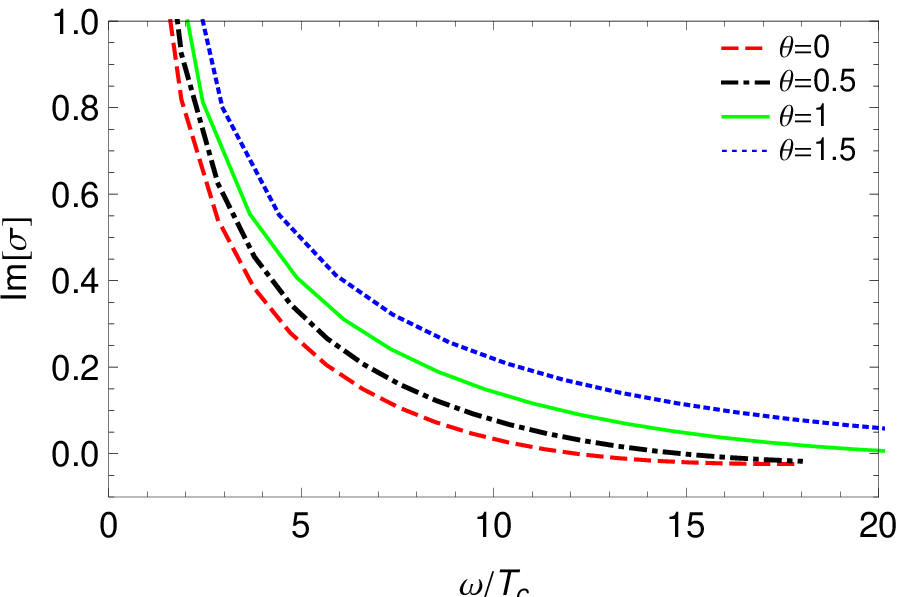}} \qquad %
\subfigure[~$\theta=0$]{\includegraphics[width=0.4\textwidth]{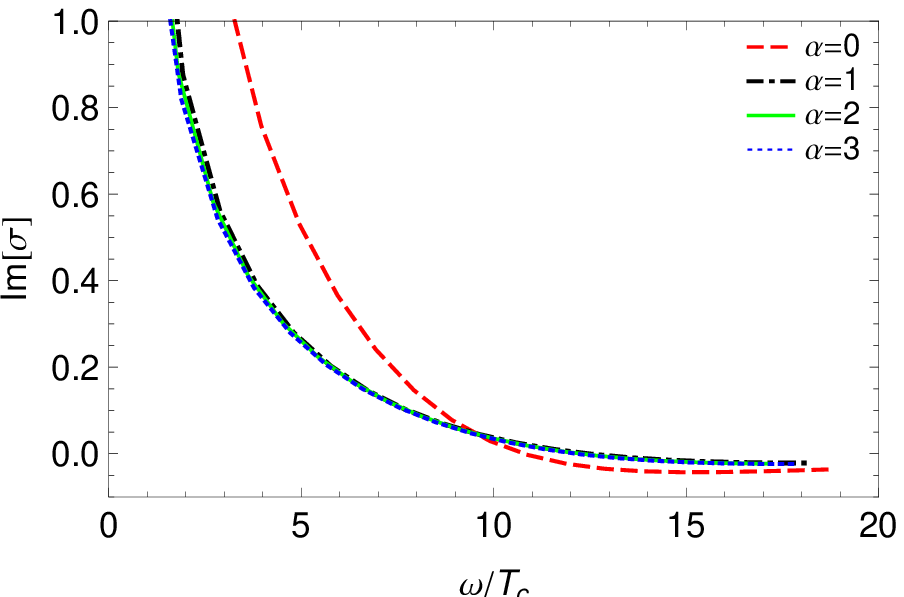}} \qquad %
\subfigure[~$\theta=1.5$]{\includegraphics[width=0.4\textwidth]{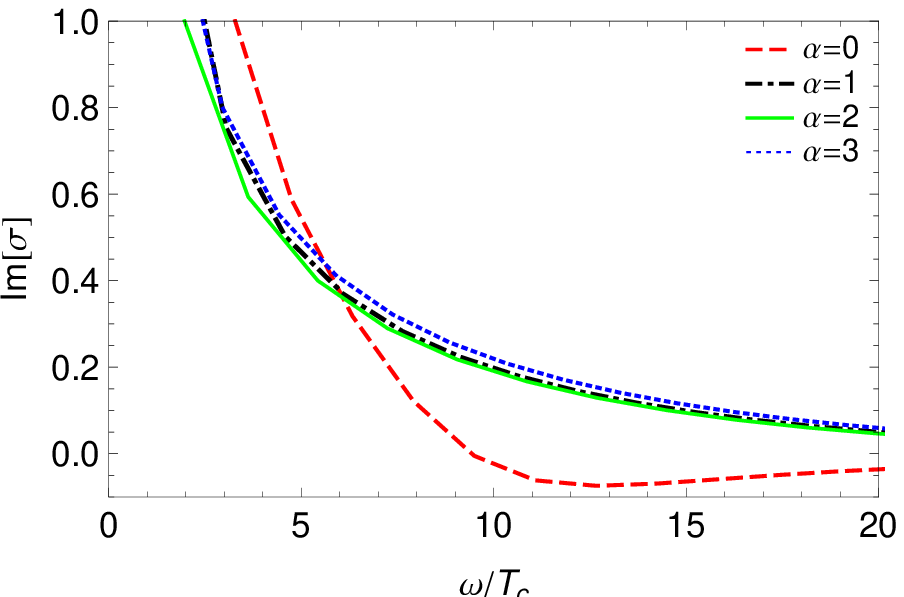}} \qquad %
\caption{The behavior of imaginary part of conductivity in $T=0.3T_c$.}\label{fig5.1}
\end{figure*}

Now, we want to explore the conductivity of holographic
superconductors with the hyperscaling violation when gauge and
scalar fields couple to each other non-minimally. So, the equation
of motion for $A_x$, due turning on a perturbation same as in the
previous section, is given by
\begin{equation}
A_{x}''+ \left[\frac{f'}{f}+\frac{2 \alpha \psi \psi '}{1+\alpha \psi ^2}+\frac{z}{r}+\frac{1}{r}\right] A_{x}'+\left[\frac{\omega ^2}{f^2 r^{2 z+2}}-\frac{2 q^2 \psi^2}{f r^{2+\theta} \left(1+\alpha \psi ^2\right)}\right]A_{x}=0,
\end{equation}
which asymptotically behaves as
\begin{equation}
A_x=A^{(0)}+\frac{A^{(1)}}{r^2},
\end{equation}
based on the AdS/CFT dictionary, the electrical conductivity equals to
\begin{equation}
\sigma=\frac{2 A^{(1)}}{i \omega A^{(0)}}.
\end{equation}
By following the numerical approach, behavior of real and
imaginary parts of conductivity as a function of $\omega/T$ are
shown in figures \ref{fig2.1} and \ref{fig3.1}. the non-minimal
coupling and the hyperscaling violation really affect the
conductivity of the holographic superconductors. However, real and
imaginary parts follow some universal behavior. For instance based
on these figures, in high frequency regimes, real and imaginary
parts of conductivity approach to a constant value based on our
choice of $\alpha$ and $\theta$. Near the zero frequency region, a
delta function is contained in the real part of the conductivity
as $Re[\sigma(\omega)]\sim\pi \delta(\omega)$ which indicates
infinite DC conductivity. While there exists a pole in the
imaginary part and it diverges $Im[\sigma(\omega)]\sim1/\omega$.
At low frequency region, real and imaginary parts are connected to
each other via Kramers-Kronig relation. In addition, for all cases
considered here according to figures \ref{fig4.1} and
\ref{fig5.1}, we observe that the gap frequency $\omega_g$ shifts
to higher frequencies when we increase the value of $\theta$ and
$\alpha$ which means increasing the effect of the hyperscaling
violation and non-minimal coupling changes the universal relation
in the gap frequency $\omega_g\sim8T_c$. larger values of $\alpha
$ and $\theta$ make the gap frequency flatter.

\section{Conclusion and Remarks}\label{sec3}
First of all, we studied holographic superconductors in the EMS
model by considering different types of non-minimal coupling of
scalar and gauge fields including an exponential, a hyperbolic
(cosh), a power-law and a fractional coupling forms. We solved the
field equations numerically with the help of shooting method and
disclose the effects of different values of the non-minimal
coupling $\alpha$ on the critical temperature and condensation for
each form of $h(\psi)$. The effects of different values of
$\alpha$ on critical temperature and condensation is not so
obvious and we obtain two groups of answers. It's reasonable
because they have the same Taylor expansion for $\alpha
\psi^{2}\ll1$. According to our results, we have superconductors
in higher temperatures for group 2 by increasing the values of
$\alpha$. While we see completely different behavior in group 1
for stronger effects of $\alpha$ and conductor/superconductor
phase transition will face with more difficulty to occur. In
addition, we have problem in numerical code in $\alpha=0.4$ for
lower temperatures in group 1. The condensation values follow the
same trend as critical temperature by increasing the strength of
non-minimal coupling. Moreover, we found out that different
choices for $\alpha$ and $h(\psi)$ do not change the order of
phase transition. In addition, we analyze the electrical
conductivity by turning on a suitable perturbation on black hole
background which corresponds to electrical current in CFT part.
While various coupling functions follow different behavior for
conductivity, almost all figures show some universal trends. For
instance in $\omega\rightarrow0$ limit, the real parts of
conductivity show a delta trend while the imaginary parts have a
pole and these two parts are connected to each other via
Kramers-Kronig relation. However at high frequency regime, both
parts tend to a constant value. Furthermore, a gap frequency forms
at low temperatures at about $\omega_g\sim8 T_c$ which shifts to
higher values by increasing the effect of non-minimal coupling.
Therefore, we understood that certain properties of
superconductors like DC conductivity and $U(1)$ spontaneous
symmetry breaking are universal and do not depend on the model.
Due to numerical calculation limitations, we can not study the
behavior of conductivity at low temperatures in some forms of
$h(\psi)$. In order to sum up, we should say that among all forms
of non-minimal coupling functions that we studied in this work,
$h(\psi)=1+\alpha\psi^2$ gives us more information about different
values of $\alpha$ at a wide range of temperature.

Secondly, from the perspective of application to realistic
systems, it is of certain interests to explore the effects of
dynamical exponent and hyperscaling violation on the holographic
superconductors. Therefore, we also have considered a holographic
superconductor model in EMS theory with a hyperscaling violating
term. At the first step, we studied the effect of different
choices of non-minimal coupling constant $\alpha$ and hyperscaling
violation term $\theta$ on condensation and critical temperature
$T_c$ numerically. Based on our results, increasing the effect of
$\alpha$ or $\theta$ makes conductor/superconductor phase
transition easier to form by increasing temperature and decreasing
condensation. Due to square root behavior of condensate, the
critical exponent of the system takes the mean-field value 1/2 in
all cases. In the next step, we analyze the electrical
conductivity in holographic setup by applying an appropriate
perturbation on the black hole background. Our choice of $\alpha$
and $\theta$ have a great impact on the shape of real and
imaginary parts of conductivity. However, they follow some
universal approaches, too.


\end{document}